\documentclass[11pt]{article}

\usepackage[T1]{fontenc}
\usepackage[margin=1in]{geometry}
\usepackage{amsmath,amssymb,amsthm,mathtools}
\numberwithin{equation}{section}
\usepackage{iftex}
\usepackage{libertinus}
\ifPDFTeX
  \usepackage{libertinust1math}
\fi
\usepackage{enumitem}
\usepackage{microtype}
\usepackage{titlesec}
\usepackage{needspace}
\usepackage{xcolor}
\usepackage{graphicx}
\usepackage[skins,breakable]{tcolorbox}
\usepackage[authoryear,round]{natbib}

\definecolor{ReferenceMulberry}{HTML}{8C3F73}
\definecolor{CitationPine}{HTML}{23644E}
\definecolor{LinkSlate}{HTML}{526776}

\usepackage{hyperref}
\hypersetup{
  colorlinks=true,
  linkcolor=ReferenceMulberry,
  citecolor=CitationPine,
  urlcolor=LinkSlate,
  linktoc=all,
  pdftitle={Second-Best Trade Is Half Efficient in Multidimensional Markets},
  pdfauthor={Dirk Bergemann, Yang Cai, Kangning Wang, Jinzhao Wu, and Renzhe Xu},
  pdfsubject={Gains from trade in multidimensional two-sided markets},
  pdfkeywords={bilateral trade, mechanism design, gains from trade, Bayesian incentive compatibility},
  pdflang={en-US}
}
\usepackage[capitalise,nameinlink,noabbrev]{cleveref}

\newcommand{\doi}[1]{\href{https://doi.org/#1}{\nolinkurl{doi:#1}}}

\newcommand{\statementtitle}[1]{\parbox[t]{\linewidth}{\setlength{\parskip}{0pt}\raggedright \let\statementcitep\citep \def\citep{\normalfont\normalsize\statementcitep}\normalfont\sffamily\fontsize{11.5}{13.5}\selectfont #1\par \vspace{4pt}}}
\newcommand{\statementheading}[1]{\statementtitle{#1}}
\newtheoremstyle{modernstatement} {0pt}{0pt}{\normalfont}{0pt}{\normalfont}{}{\newline} {\statementheading{{\bfseries #1~#2}\if\relax\detokenize{#3}\relax\else\enspace\textemdash\enspace #3\fi}}
\theoremstyle{modernstatement}
\newtheorem{theorem}{Theorem}[section]
\newtheorem{proposition}[theorem]{Proposition}
\newtheorem{lemma}[theorem]{Lemma}

\newtheorem{definition}[theorem]{Definition}
\newtheorem{example}[theorem]{Example}
\newtheorem{remark}[theorem]{Remark}

\tcbset{statementbox/.style={
  enhanced, unbreakable, arc=12pt, outer arc=12pt,
  colback=white, colframe=black!50, boxrule=0.75pt,
  left=15pt, right=15pt, top=11pt, bottom=11pt, boxsep=0pt,
  before skip=13pt plus 2pt minus 2pt,
  after skip=12pt plus 2pt minus 2pt
}}
\tcolorboxenvironment{theorem}{statementbox}
\tcolorboxenvironment{proposition}{statementbox}
\tcolorboxenvironment{lemma}{statementbox}
\tcolorboxenvironment{corollary}{statementbox}
\tcolorboxenvironment{definition}{statementbox}
\tcolorboxenvironment{example}{statementbox}
\tcolorboxenvironment{remark}{statementbox}
\newenvironment{restatedstatement}[3]{\begin{tcolorbox}[statementbox] \statementheading{{\bfseries\hyperref[#2]{#1~\ref*{#2}}}\enspace\textemdash\enspace #3\enspace(restated)}\par \nobreak }{\end{tcolorbox}}
\newtcolorbox{programbox}{
  statementbox, breakable, lines before break=6,
  skin first=enhanced, skin middle=enhanced, skin last=enhanced,
  toprule at break=0.75pt, bottomrule at break=0.75pt,
  left=8pt, right=8pt,
  colbacktitle=white, coltitle=black,
  fonttitle=\normalfont, titlerule=0pt, toptitle=11pt, bottomtitle=0pt,
  title after break={\statementtitle{{\bfseries\hyperref[prog:second-best]{Program~(\theprogram)}}\enspace\textemdash\enspace Second-best program\enspace(continued)}}
}

\crefname{theorem}{Theorem}{Theorems}
\Crefname{theorem}{Theorem}{Theorems}
\crefname{proposition}{Proposition}{Propositions}
\Crefname{proposition}{Proposition}{Propositions}
\crefname{lemma}{Lemma}{Lemmas}
\Crefname{lemma}{Lemma}{Lemmas}
\crefname{corollary}{Corollary}{Corollaries}
\Crefname{corollary}{Corollary}{Corollaries}
\crefname{definition}{Definition}{Definitions}
\Crefname{definition}{Definition}{Definitions}
\crefname{example}{Example}{Examples}
\Crefname{example}{Example}{Examples}
\crefname{remark}{Remark}{Remarks}
\Crefname{remark}{Remark}{Remarks}
\crefname{section}{Section}{Sections}
\Crefname{section}{Section}{Sections}
\crefname{subsection}{Section}{Sections}
\Crefname{subsection}{Section}{Sections}
\crefname{appendix}{Appendix}{Appendices}
\Crefname{appendix}{Appendix}{Appendices}
\crefname{equation}{}{}
\Crefname{equation}{}{}
\newcounter{program}
\renewcommand{\theprogram}{P}
\crefname{program}{Program}{Programs}
\Crefname{program}{Program}{Programs}
\crefformat{program}{#2Program~(#1)#3}
\Crefformat{program}{#2Program~(#1)#3}
\newcommand{\programsymbol}{\hyperref[prog:second-best]{\ensuremath{\mathrm{P}}}}
\crefformat{equation}{#2(#1)#3}
\Crefformat{equation}{#2(#1)#3}
\crefrangeformat{equation}{#3(#1)#4--#5(#2)#6}
\Crefrangeformat{equation}{#3(#1)#4--#5(#2)#6}
\crefmultiformat{equation}{#2(#1)#3}{ and~#2(#1)#3}{, #2(#1)#3}{ and~#2(#1)#3}
\Crefmultiformat{equation}{#2(#1)#3}{ and~#2(#1)#3}{, #2(#1)#3}{ and~#2(#1)#3}
\crefrangemultiformat{equation}{#3(#1)#4--#5(#2)#6}{ and~#3(#1)#4--#5(#2)#6}{, #3(#1)#4--#5(#2)#6}{ and~#3(#1)#4--#5(#2)#6}
\Crefrangemultiformat{equation}{#3(#1)#4--#5(#2)#6}{ and~#3(#1)#4--#5(#2)#6}{, #3(#1)#4--#5(#2)#6}{ and~#3(#1)#4--#5(#2)#6}
\newcommand{\namedstatementref}[4]{\texorpdfstring{\hyperref[#3]{\if\relax\detokenize{#2}\relax\nameref*{#3}\else#2\fi\space(\Cref*{#3}#4)}}{\if\relax\detokenize{#2}\relax\getrefbykeydefault{#3}{name}{}\else#2\fi\space(#1 \getrefnumber{#3}#4)}}
\newcommand{\thmref}[2][]{\namedstatementref{Theorem}{#1}{#2}{}}
\newcommand{\lemref}[2][]{\namedstatementref{Lemma}{#1}{#2}{}}
\newcommand{\lempartref}[3][]{\namedstatementref{Lemma}{#1}{#2}{#3}}
\newcommand{\propref}[2][]{\namedstatementref{Proposition}{#1}{#2}{}}
\newcommand{\exampleref}[2][]{\namedstatementref{Example}{#1}{#2}{}}
\makeatletter
\newcommand{\namedstatementlabel}[3]{\def\@currentlabelname{#3}\label[#1]{#2}}
\makeatother

\allowdisplaybreaks
\titlespacing*{\section}{0pt}{2.4ex plus 0.6ex minus 0.3ex}{1.0ex}
\titlespacing*{\subsection}{0pt}{1.8ex plus 0.5ex minus 0.2ex}{0.7ex}
\newcommand{\paratitle}[1]{\par\noindent\textbf{#1.}\enspace\ignorespaces}
\AddToHook{cmd/subsection/before}{\Needspace{8\baselineskip}}
\setlist[enumerate]{leftmargin=2em,itemsep=0.35em,topsep=0.5em}
\setlist[itemize]{leftmargin=2em,itemsep=0.35em,topsep=0.5em}
\newcommand{\E}{\mathbb E}
\providecommand{\R}{}\renewcommand{\R}{\mathbb R}
\newcommand{\cF}{\mathcal F}
\newcommand{\cX}{\mathcal X}
\newcommand{\cA}{\mathcal A}
\newcommand{\cB}{\mathcal B}
\newcommand{\cU}{\mathcal U}
\newcommand{\cL}{\mathcal L}
\newcommand{\cN}{\mathcal N}
\newcommand{\cM}{\mathcal M}
\newcommand{\FB}{\mathrm{FB}}
\newcommand{\SB}{\mathrm{SB}}
\newcommand{\GFT}{\mathrm{GFT}}
\newcommand{\1}{\mathbf 1}
\newcommand{\Id}{\mathbb{I}}
\newcommand{\tv}{\widetilde v}
\newcommand{\tc}{\widetilde c}
\newcommand{\tH}{\widetilde H}
\providecommand{\Beta}{}\renewcommand{\Beta}{\mathrm B}
\newcommand{\thB}{\theta_B}
\newcommand{\thS}{\theta_S}
\newcommand{\Out}{\mathrm{out}}
\providecommand{\In}{}\renewcommand{\In}{\mathrm{in}}
\DeclareMathOperator{\Unif}{Unif}
\DeclareMathOperator{\Exp}{Exp}
\DeclareMathOperator{\conv}{conv}

\newcommand{\paperauthor}[3]{\begin{minipage}[t]{0.30\textwidth} \centering \setlength{\parskip}{0pt}{\fontsize{11.5}{14}\selectfont\bfseries #1\par} \vspace{2pt}{\fontsize{9.5}{12}\selectfont\color{black!70}#2\par} \vspace{1pt}{\fontsize{9}{11}\selectfont\href{mailto:#3}{\nolinkurl{#3}}\par} \end{minipage}}
\title{Second-Best Trade Is Half Efficient in\\ Multidimensional Markets}
\newcommand{\paperauthors}{\makebox[\textwidth][c]{\paperauthor{Dirk Bergemann}{Yale University}{dirk.bergemann@yale.edu}\hspace{0.015\textwidth}\paperauthor{Yang Cai}{Yale University}{yang.cai@yale.edu}\hspace{0.015\textwidth}\paperauthor{Kangning Wang}{Rutgers University}{kn.w@rutgers.edu}}\par\vspace{8pt} \paperauthor{Jinzhao Wu}{Yale University}{jinzhao.wu@yale.edu}\hspace{0.015\textwidth}\paperauthor{Renzhe Xu}{Shanghai University of\\Finance and Economics}{xurenzhe@sufe.edu.cn}}
\author{\paperauthors}
\date{}

\makeatletter
\renewcommand{\maketitle}{\begin{center} \setlength{\parskip}{0pt}\vspace*{4pt}{\LARGE\bfseries\@title\par} \vspace{22pt}\@author\par \vspace{15pt}{\color{CitationPine!55}\rule{0.88\textwidth}{0.6pt}\par} \end{center} \vspace{-15pt}}
\makeatother

\begin{document}
\hypersetup{pageanchor=false}
\pagenumbering{gobble}
\maketitle
\thispagestyle{empty}
\begin{abstract}
We prove that at least half of the first-best gains from trade can always be achieved in multidimensional two-sided markets with additive values and costs and arbitrary downward-closed feasibility constraints. The guarantee is achieved by a mechanism satisfying Bayesian incentive compatibility, interim individual rationality, and strong budget balance, and is tight already in single-item bilateral trade. Types are independent across agents, while the coordinates of each agent's type may be arbitrarily correlated. Feasibility constraints may couple trades across items and agents. This extends the known tight one-half
guarantee for single-item bilateral trade.

Our proof develops a new application of the
Cai--Devanur--Weinberg duality framework. We show that the
objective of every feasible dual is at least half of the first
best. Strong duality then yields the guarantee without
characterizing an optimal mechanism or dual. The key observation
is that every feasible dual induces continuous-time Markov
chains that, when stopped by exponential clocks, recover true
types as expectations of virtual types. The dual constraints
control the rate at which these chains change the prior
distributions. We establish our bound by combining this control
with a coupling of the agents' clocks.
\end{abstract}

\par\vspace{-0.75em}
\noindent\makebox[\textwidth][c]{\color{CitationPine!55}\rule{0.88\textwidth}{0.6pt}}\par
\vspace{0em}
\begingroup
\makeatletter
\renewcommand*\l@section[2]{%
  \ifnum \c@tocdepth >\z@
    \addpenalty\@secpenalty
    \addvspace{0.1em \@plus\p@}%
    \setlength\@tempdima{1.5em}%
    \begingroup
      \parindent \z@ \rightskip \@pnumwidth
      \parfillskip -\@pnumwidth
      \leavevmode \bfseries
      \advance\leftskip\@tempdima
      \hskip -\leftskip
      #1\nobreak\hfil \nobreak\hb@xt@\@pnumwidth{\hss #2}\par
    \endgroup
  \fi}
\makeatother
\fontsize{9}{11}\selectfont
\setlength{\parskip}{0pt}
\hypersetup{linkcolor=LinkSlate}
\tableofcontents
\endgroup
\thispagestyle{empty}
\clearpage
\pagenumbering{arabic}
\hypersetup{pageanchor=true}
\section{Introduction}
\label{sec:model}
Facilitating mutually beneficial trade is a fundamental economic problem.
We study this problem in two-sided markets, where buyers have private
values for goods or services and sellers have private costs of providing
them. For example, on short-term rental platforms, guests value stays
at different properties, while hosts incur costs from making their
properties available. The goal is to maximize gains from trade---the
total value generated for buyers minus the total cost incurred by sellers.

The \emph{first best} is the maximum expected gains from trade achievable
subject only to feasibility. The seminal impossibility theorem of
\citet*{MyersonSatterthwaite1983} shows that, in general, no Bayesian
incentive-compatible, interim individually rational, and budget-balanced
mechanism can attain this benchmark, even in bilateral trade with one
buyer, one seller, and one item. The \emph{second best} is the largest
expected gains from trade achievable under these constraints.
This raises an important question: what fraction of the first-best
gains from trade can the second best always guarantee?

In a breakthrough, \citet*{DengMaoSivanWang2022} gave the first
constant-factor approximation to the first best in bilateral trade,
achieving a $1/8.23$ fraction. \citet*{Fei2022} improved this guarantee
to $1/3.15$, and \citet*{HartlineWang2025} gave a geometric proof of
the same bound. More recently, \citet*{LiuQinRenWang2026} settled
the question for bilateral trade by proving that the second best
is at least half of the first best and that this bound is tight.
\begin{quote}
\begin{center}
\itshape
What fraction of the first-best gains from trade can the second best\\
always guarantee in \textbf{multidimensional} two-sided markets?
\end{center}
\end{quote}
We prove that the second best is at least half of the first best
in a general class of multidimensional two-sided markets, and
that this guarantee is tight. A central conceptual obstacle is
the lack of a useful structural characterization of the
second-best mechanism in this setting. We overcome this obstacle
through a new application of the Cai--Devanur--Weinberg duality
framework~\citep*{CaiDevanurWeinberg2021}, connecting arbitrary
feasible dual solutions to continuous-time Markov chains whose
evolution can be controlled using the dual constraints. We use this connection to prove that the objective of every
feasible dual is at least half of the first best. Strong duality identifies the second best with the minimum of
these objectives, yielding the guarantee without characterizing
an optimal mechanism or dual. The connection may also be of independent
interest.

We first describe the model and state our main result.
A proof overview appears in \Cref{sec:overview}.

\subsection{Model: trades, types, and feasibility}
\label{sec:model-trades}

A \emph{multidimensional two-sided market} is specified by a tuple
\[
  \cN=\big(I,J,E,i(\cdot),j(\cdot),\cF,
  (F_i)_{i\in I},(G_j)_{j\in J}\big),
\]
where $I$ and $J$ are finite disjoint sets of buyers and sellers,
and $E$ is a finite set of possible trades. Each trade $e\in E$
involves a buyer $i(e)\in I$ and a seller $j(e)\in J$; multiple
distinct trades may involve the same buyer--seller pair. We write
\[
  E_i:=\{e\in E:i(e)=i\},\qquad
  E_j:=\{e\in E:j(e)=j\}
\]
for the sets of trades incident to buyer $i$ and seller $j$,
respectively.

Buyer $i$ has a private type $v_i=(v_{i,e})_{e\in E_i}$ drawn
from a distribution $F_i$ on $\R_+^{E_i}$, and seller $j$ has
a private type $c_j=(c_{j,e})_{e\in E_j}$ drawn from a
distribution $G_j$ on $\R_+^{E_j}$. Here $v_{i,e}$ is the value
buyer $i$ derives from executing trade $e$, and $c_{j,e}$ is
the cost seller $j$ incurs. Values and costs are additive
across executed trades.

Let $T_i$ and $T_j$ denote the finite nonempty supports of
$F_i$ and $G_j$, and let $f_i$ and $g_j$ be their probability
mass functions, strictly positive on these supports.
Coordinates within an agent's type may be arbitrarily correlated,
while types are independent across agents. We write
\[
  T_I:=\prod_{i\in I}T_i,\qquad
  T_J:=\prod_{j\in J}T_j
\]
for the buyer and seller type-profile spaces.

A publicly known \emph{feasible family} $\cF\subseteq 2^E$
specifies the sets of trades that may be executed. We assume
that $\cF$ is nonempty and \emph{downward closed}: if
$X\in\cF$ and $X'\subseteq X$, then $X'\in\cF$.
A randomized allocation is a distribution over $\cF$.
Its vector $x$ of marginal execution probabilities belongs
to the polytope
\[
  \cX(\cF):=\conv\{\1_X:X\in\cF\}\subseteq[0,1]^E,
\]
where $\1_X$ is the incidence vector of $X$. For fixed types,
additivity implies that expected values and costs depend only
on these marginals. This model captures \exampleref[single-item bilateral trade]{ex:bilateral},
\exampleref[matching markets with scalar private types]{ex:matching-model}, \exampleref[markets with one constrained-additive buyer and heterogeneous items]{ex:multi}, and~\exampleref[bilateral bargaining]{ex:bargaining} as special cases (\Cref{sec:examples}). Formal definitions of
mechanisms, incentive compatibility, individual rationality, budget
balance, and the first-best and second-best benchmarks are given in
\Cref{sec:model-mechanisms}.

\subsection{Main result}
\label{sec:main}

\begin{theorem}[Main theorem]\label[theorem]{thm:main}
For every multidimensional two-sided market $\cN$, there exists
a direct mechanism satisfying Bayesian incentive compatibility,
interim individual rationality, and strong budget balance that
achieves at least half of the first-best gains from trade. Equivalently,
\[
  \SB\ge\tfrac12\,\FB .
\]
\end{theorem}

\begin{remark}[The constant is tight]\label[remark]{rem:tightness}
The constant one half in the \thmref[main theorem]{thm:main} is tight, already in bilateral trade: \citet*[Theorem~6.2]{LiuQinRenWang2026} show that no Bayesian incentive-compatible, interim individually rational, and weakly budget-balanced mechanism for bilateral trade guarantees more than half of the first best.
\end{remark}

\section{Related work}
\label{sec:related}

\paratitle{Concurrent results}
We discuss two works developed independently and concurrently with ours.
\citet*{BeiLiWuZhou2026} establish the sharp one-half guarantee for
single-dimensional matching markets with arbitrary downward-closed
feasibility constraints. Applying their result separately to each
item also gives a one-half guarantee for additive multidimensional
markets whose feasibility constraints separate across items.
Using a reduction of \citet*{CaiGoldnerMaZhao2021}, they additionally
obtain a one-quarter guarantee for a single unit-demand buyer with
mutually independent item values and seller costs under
dominant-strategy incentive compatibility and ex-post individual
rationality. For this market structure, our theorem gives a
one-half guarantee under the weaker requirements of Bayesian
incentive compatibility and interim individual rationality,
while allowing arbitrary correlation among the buyer's item values.

\citet*{LiuQinWang2026} develop a reduction that transfers universal
bilateral-trade guarantees to single-dimensional matching markets
with downward-closed feasibility without additional loss.
Applied to the result of \citet*{LiuQinRenWang2026}, their reduction
yields the one-half guarantee for these markets. It also yields
sharper guarantees under additional distributional restrictions.

Our theorem establishes the universal one-half guarantee for
multidimensional private types under arbitrary downward-closed
allocation constraints, including constraints that couple trades
across items and agents. Both concurrent works exploit the
structure of single-dimensional mechanism design through payment
identities and interim monotonicity. A comparable structural
characterization of the second-best mechanism is unavailable in
our multidimensional setting, where even one-sided optimal
mechanisms can be highly complex
\citep*{HartReny2015,HartNisan2019,DaskalakisDeckelbaumTzamos2017}.
Our proof establishes the guarantee without characterizing an
optimal mechanism or dual. By dualizing incentive, participation,
and budget constraints, we show that the objective of every
feasible dual is at least half of the first best
(\Cref{sec:overview}).

\paratitle{Offering mechanisms}
Another line of work analyzes offering mechanisms, beginning with
random offering in bilateral trade, where a randomly chosen side
makes a take-it-or-leave-it offer
\citep*{DengMaoSivanWang2022,Fei2022,HartlineWang2025}.
For single-dimensional matching markets,
\citet*{BabaioffRubinsteinTanWang2026} obtain a $1/3.15$ fraction
of the first best through generalized offering mechanisms.
\citet*{BeiWuZhou2026} improve this guarantee to $1/e$ and show
that it is tight for the generalized random-offerer mechanism
over downward-closed matching markets.

With multidimensional types, the proposer's optimal one-sided
mechanism generally lacks a tractable characterization.
Nevertheless, explicit mechanisms can provide useful lower
bounds on the proposer's optimal profit
\citep*{CaiGoldnerMaZhao2021,RubinsteinTanZhou2026}.
In particular, \citet*{RubinsteinTanZhou2026} obtain constant-factor
guarantees relative to the first best for one additive seller and
one or more XOS buyers under item-independence assumptions on
values and costs.

Approximating optimal one-sided profit becomes substantially
harder under arbitrary within-agent correlation: even for a single
additive buyer with zero seller costs, no uniformly bounded menu
size guarantees a constant fraction of optimal revenue
\citep*{HartNisan2019}. Our proof avoids this intermediate
revenue-approximation problem. For additive values and costs,
it establishes the tight one-half guarantee under arbitrary
within-agent correlation and general downward-closed allocation
constraints.
\section{Preliminaries}
\label{sec:model-mechanisms}
A \emph{direct mechanism} $M=(x,p)$ consists of an allocation rule
$x:T_I\times T_J\to\cX(\cF)$ and payment rules (\emph{transfers})
$p=\big((p_i)_{i\in I},(p_j)_{j\in J}\big)$, comprising buyer payments
$p_i:T_I\times T_J\to\R_+$ and seller receipts
$p_j:T_I\times T_J\to\R_+$. At a report profile $(v,c)$,
trade $e$ is executed with probability $x_e(v,c)$, buyer $i$
pays $p_i(v,c)$, and seller $j$ receives $p_j(v,c)$.

Agents are risk neutral and have quasilinear utilities.
Under truthful reporting, their expected utilities, averaged
over the mechanism's allocation randomness, are
\begin{align*}
  u_i(v,c)&:=v_i\cdot x_{E_i}(v,c)-p_i(v,c),\\
  u_j(v,c)&:=p_j(v,c)-c_j\cdot x_{E_j}(v,c),
\end{align*}
where $x_{E_i}(v,c)$ and $x_{E_j}(v,c)$ denote the restrictions
of the allocation vector to the trades incident to buyer $i$
and seller $j$, respectively.

The \emph{interim utility} of buyer $i$ with true type $v_i$
who reports $v_i'$ is
\[
  \cU_i(v_i;v_i')
  :=\E_{v_{-i},c}\big[
    v_i\cdot x_{E_i}(v_i',v_{-i},c)
    -p_i(v_i',v_{-i},c)
  \big].
\]
The expectation is over the other agents' types: the allocation
and payment are evaluated at the reported type $v_i'$, while
the allocation is valued using the true type $v_i$. In particular,
$\cU_i(v_i;v_i)=\E_{v_{-i},c}\big[u_i(v_i,v_{-i},c)\big]$.

Similarly, seller $j$'s interim utility with true type $c_j$
and report $c_j'$ is
\[
  \cU_j(c_j;c_j')
  :=\E_{v,c_{-j}}\big[
    p_j(v,c_j',c_{-j})
    -c_j\cdot x_{E_j}(v,c_j',c_{-j})
  \big].
\]

\Needspace{12\baselineskip}
We consider mechanisms that are \emph{Bayesian incentive compatible}
(BIC), \emph{interim individually rational} (interim IR), and
\emph{strongly budget balanced at every report profile}
(reportwise SBB):
\begin{gather}
  \cU_i(v_i;v_i)\ge\cU_i(v_i;v_i'),\qquad
  \cU_j(c_j;c_j)\ge\cU_j(c_j;c_j'),
  \label{eq:bic}\\
  \cU_i(v_i;v_i)\ge0,\qquad
  \cU_j(c_j;c_j)\ge0,
  \label{eq:ir}\\
  \sum_{i\in I}p_i(v,c)=\sum_{j\in J}p_j(v,c).
  \label{eq:sbb}
\end{gather}
These conditions hold for all agents $i\in I$ and $j\in J$,
all types $v_i,v_i'\in T_i$ and $c_j,c_j'\in T_j$, and all
report profiles $(v,c)\in T_I\times T_J$.
Let $\cM$ denote the class of direct mechanisms satisfying
\Cref{eq:bic,eq:ir,eq:sbb}.

Executing trade $e$ generates the gain
$v_{i(e),e}-c_{j(e),e}$. The \emph{expected gains from trade}
of a mechanism $M=(x,p)$ are therefore
\begin{equation}\label{eq:welfare}
  \GFT(x)
  :=\E_{v,c}\Big[
    \sum_{e\in E}x_e(v,c)
    \big(v_{i(e),e}-c_{j(e),e}\big)
  \Big].
\end{equation}

We compare two benchmarks. Define the maximum feasible surplus
at a type profile by
\begin{equation}\label{eq:surplus}
  H(v,c)
  :=\max_{X\in\cF}
    \sum_{e\in X}\big(v_{i(e),e}-c_{j(e),e}\big).
\end{equation}
Since $\varnothing\in\cF$, we have $H(v,c)\ge0$.
The \emph{first best} is the maximum expected gains from trade
subject only to feasibility:
\[
  \FB:=\E_{v,c}\big[H(v,c)\big].
\]
The \emph{second best} additionally imposes Bayesian incentive
compatibility, interim individual rationality, and reportwise
strong budget balance:
\[
  \SB:=\sup\big\{\GFT(x):M=(x,p)\in\cM\big\}.
\]
\section{Proof overview}
\label{sec:overview}
\paratitle{Approximating the first best using any dual}
Our approach uses the Lagrangian duality framework of
\citet*{CaiDevanurWeinberg2021}, extended to two-sided markets by
\citet*{BrustleCaiWuZhao2017}. A common use of this framework is
to fix a convenient dual solution, whose objective upper-bounds
the performance of every mechanism, and then design a mechanism
that approximates this benchmark
\citep*{CaiZhao2017,CaiGoldnerMaZhao2021}.
We instead prove that the objective of \emph{every} feasible dual
is at least half of the first best. Strong duality identifies the
second best with the minimum of these objectives, yielding the
desired guarantee. Our argument uses only properties shared by
all feasible duals, allowing us to handle multidimensional types
without characterizing an optimal dual.

\paratitle{From virtual values back to true values}
The dual is obtained by attaching multipliers to the incentive,
participation, and budget constraints of the second-best linear program
(\Cref{sec:program}). The incentive and participation multipliers form a
flow on each agent's type space, while a scalar $\tau\ge0$ multiplies the
budget constraint. Fix an arbitrary dual. Its flows transform true
values and costs into normalized virtual values and costs, and by
\thmref[duality]{prop:dual} its objective is
\[
    D=(1+\tau)\E[\tH(v,c)],
\]
where $\tH$ is the maximum feasible surplus evaluated at these virtual
values and costs. The first best is the same
surplus maximization evaluated at true values and costs. We will prove
$\FB\le2D$ for every dual.

\paratitle{Step 1: a Markov-chain inverse}
For each agent, the linear transformation from true types to virtual types can be expressed as
$\Id-\cA$, where $\cA$ is a Markov generator on the agent's type
space that depends on the chosen dual: its off-diagonal entries are nonnegative and its rows sum to zero.
The same matrix acts on every coordinate of the agent's type. The first key step is to rewrite the inverse of $\Id-\cA$ using a probabilistic representation that we refer to as the \lemref[\emph{Markov recovery process}]{lem:markov}:
\[
    (\Id-A)^{-1}=\int_0^\infty e^{-\theta}e^{\theta \cA}\,\mathrm{d}\theta.
\]
The process runs the continuous-time
chain with generator $A$ from the true type for an independent time
$\theta\sim\Exp(1)$.\footnote{The transition matrix of the Markov chain after time $\theta$ is $e^{\theta \cA}$.}  The true type is therefore
the expected virtual type evaluated at the terminal state.
The familiar single-dimensional identity
$\E[\varphi(t)\mid t\ge v]=v$, where $t$ is drawn from the prior and
$\varphi$ is Myerson's virtual value, arises as a \emph{limiting} special
case.\footnote{For the canonical choice of dual variables, i.e., the downward flow (\Cref{sec:program}), as
$\tau\to\infty$, the normalized virtual value converges to $\varphi$,
and the terminal-type distribution converges to the prior conditioned
on being at least $v$.}

Maximum feasible surplus is a convex function of the vector of trade
gains. Applying Jensen's inequality to the recovery identity therefore gives
\[
    \FB\le\E[\tH(\hat v,\hat c)],
\]
where $\hat v$ and $\hat c$ are the terminal type profiles.
This inequality holds for any coupling of the terminal types that
preserves each agent's recovery distribution conditional on the
initial type profile. Thus, although the initial types are drawn
independently, their terminal types need not be independent.
We exploit this freedom to choose a coupling under which expected
virtual surplus at the terminal profile can be bounded by its
expectation under the prior, up to a controlled factor in the next step.

\paratitle{Step 2: density bounds
and coupled clocks}
We first control the terminal distribution conditional on a fixed clock
value, and then choose a coupling of the clocks to obtain the desired
comparison with the prior.

For a single agent whose initial type is drawn from $f$, the terminal
distribution conditional on the clock taking value $\theta$ is $p_\theta:=f e^{\theta \cA}$.
Flow conservation, together with the normalization of $\cA$, gives
$f\cA\le\kappa f$ componentwise, where $\kappa:=\tau/(1+\tau)$.
Since $e^{\theta \cA}$ has nonnegative entries,
\[
    \frac{\mathrm{d}}{\mathrm{d}\theta}p_\theta
    =f\cA e^{\theta \cA}
    \le\kappa f e^{\theta \cA}
    =\kappa p_\theta.
\]
Thus, each type's probability increases at a rate of at most $\kappa$
times its current probability. Integrating this inequality gives the
componentwise \lemref[prior-density bound]{lem:density},
$p_\theta\le e^{\kappa\theta}f$.

\begin{figure}[!t]
    \centering
    \includegraphics[width=.83\linewidth]
        {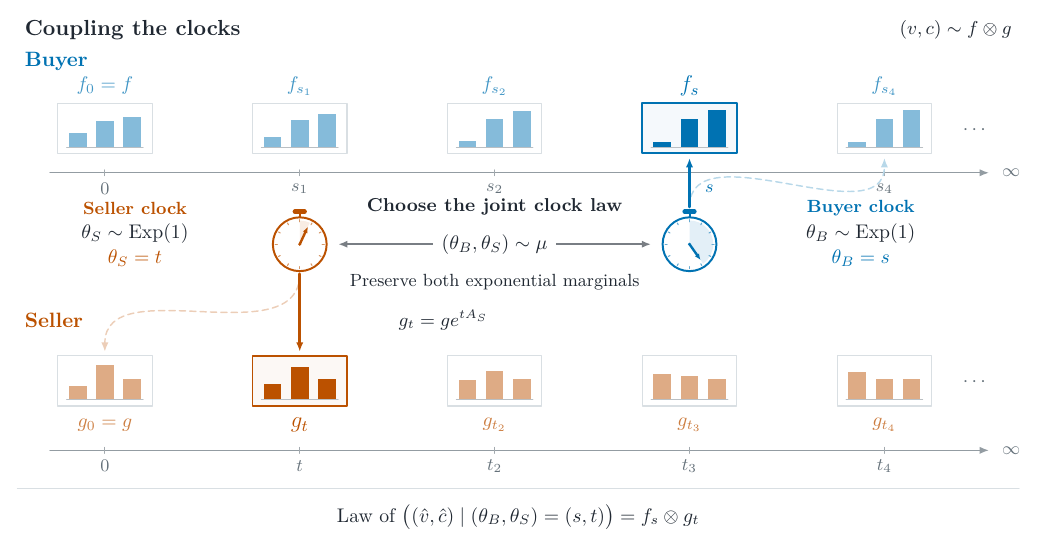}
    \caption{
    Coupling the clocks of the Markov recovery processes.
    The two rows show distributions at successive times for chains
    started independently from the buyer and seller priors.
    We choose a joint clock law $\mu$ with
    $\Exp(1)$ marginals, independently of the initial
    types and chain randomness. The two chains use independent
    randomness.
    Solid arrows show one paired realization $(s,t)$;
    dashed arrows indicate other possible stopping times.
    Conditional on the selected times, the terminal distribution
    is $f_s\otimes g_t$.
    }
    \label{fig:clock-coupling}
\end{figure}

We now use this bound to choose a coupling of the recovery processes. In bilateral trade, draw the clocks $(\thB,\thS)$ from any joint law $\mu$ with $\Exp(1)$ marginals, independently of the types and the chains, as illustrated in~\Cref{fig:clock-coupling}. Given $(\thB,\thS)=(s,t)$, the terminal types are independent, with laws $f_s:=fe^{s \cA_B}\le e^{\kappa s}f$ and $g_t:=ge^{t \cA_S}\le e^{\kappa t}g$, where $\cA_B$ and $\cA_S$ are the buyer's and seller's generators. Since $\tH\ge0$,
\[
  \E\big[\tH(\hat v,\hat c)\,\big|\,\thB=s,\ \thS=t\big]\le e^{\kappa(s+t)}\,\E\big[\tH(v,c)\big].
\]

Averaging over $\mu$ and combining with the prior-density bounds gives $\FB\le\E_\mu[e^{\kappa(\thB+\thS)}]\cdot D/(1+\tau)$. Independent clocks give $\E_\mu[e^{\kappa(\thB+\thS)}]=(1+\tau)^2$, and hence only $\FB\le(1+\tau)D$.

We instead couple the clocks antithetically, $\thB=-\log u$ and $\thS=-\log(1-u)$ for a single $u\sim\Unif(0,1)$: both marginals remain $\Exp(1)$, but a long buyer clock is always paired with a short seller clock. Then
\[
  \E_\mu\big[e^{\kappa(\thB+\thS)}\big]=\int_0^1\big[u(1-u)\big]^{-\kappa}\,\mathrm{d}u<2(1+\tau)
\]
by the \lemref[beta-integral bound]{lem:beta}, so $\FB\le2D$ in bilateral trade.

\paratitle{Step 3: thinning general markets}
For a general market, multiplying density bounds over all agents
would cause an exponential blowup. Instead, fix shared clocks,
one per side, and write $K_B=e^{\kappa\thB}$ and
$K_S=e^{\kappa\thS}$. We compare the terminal profile to an
auxiliary profile drawn from the original product prior, whose
expected virtual surplus is precisely $D/(1+\tau)$.
The density bounds allow us to couple these profiles independently
across agents conditional on the clocks, so that each agent's
terminal and auxiliary types agree with probability at least
$1/K_B$ for buyers and $1/K_S$ for sellers, even conditional
on the terminal profile.

Take an allocation maximizing virtual surplus at the terminal
profile and retain only trades for which both the buyer's and
seller's terminal types agree with their corresponding auxiliary
types. The retained allocation is feasible at the auxiliary
profile and preserves the virtual gain of each retained trade.
Each selected trade has nonnegative virtual gain and survives
with conditional probability at least $1/(K_BK_S)$, so linearity
of expectation loses only \lemref[one density factor per side]{lem:one-factor}.
Averaging over the antithetic clocks gives $\FB\le2D$ for every
feasible dual; strong duality completes the proof.

\section{The duality framework and the Markov recovery process}
\label{sec:duality}

We express the second best through dual flows and virtual values, then
derive the Markov recovery identity and the prior-density bounds used
in the subsequent proofs.

\subsection{The duality framework}
\label{sec:program}

We use the Lagrangian duality framework of
\citet*{CaiDevanurWeinberg2021}, in the two-sided form of
\citet*{BrustleCaiWuZhao2017}. The LP reformulation and duality
derivations are given in \Cref{app:payment,app:dual}.

\paratitle{The second best as a linear program}
Two modifications put the second-best problem into an equivalent form
suited to dualization. First, we replace strong budget
balance and nonnegative transfers with ex-ante weak budget balance
and unrestricted transfers. Building on \citet*{BrustleCaiWuZhao2017},
we establish \propref[budget-balance equivalence]{prop:payment}:
this relaxation preserves the implementable allocation rules.
Second, we add a null report $\bot$
that gives the reporting agent zero allocation and zero transfer,
so interim individual rationality becomes the incentive constraint
against this report. Write $T_i^+:=T_i\cup\{\bot\}$ and
$T_j^+:=T_j\cup\{\bot\}$, and set $\cU_i(v_i;\bot):=0$ and
$\cU_j(c_j;\bot):=0$. We can thus express the
\lemref[second best as a linear program]{lem:sb-is-lp}:
\begin{equation}\label{eq:lp-compact}
\begin{aligned}
  \max_{x,\,p}\ \ &\GFT(x)\\
  \text{s.t.}\ \ &\cU_i(v_i;v_i)\ge\cU_i(v_i;v_i') &&[\alpha_i(v_i,v_i')] &&(i\in I,\ v_i\in T_i,\ v_i'\in T_i^+),\\
  &\cU_j(c_j;c_j)\ge\cU_j(c_j;c_j') &&[\beta_j(c_j,c_j')] &&(j\in J,\ c_j\in T_j,\ c_j'\in T_j^+),\\
  &\E_{v,c}\Big[\sum_{i\in I}p_i(v,c)-\sum_{j\in J}p_j(v,c)\Big]\ge0 &&[\tau],\\
  &x(v,c)\in\cX(\cF) &&&&\text{for every profile }(v,c),
\end{aligned}
\end{equation}
stated in full as \Cref{prog:second-best} in \Cref{app:payment}. The brackets name the nonnegative multiplier that we attach to each constraint.

\paratitle{Dual variables are flows}
We dualize every constraint except feasibility. Since transfers are
unrestricted, the dual objective is finite only when every transfer
coefficient vanishes. After factoring out the probability of the other
agents' types, buyer $i$'s payment at report $v_i$ has coefficient
$\tau f_i(v_i)$ from the budget constraint, a contribution
$-\alpha_i(v_i,v_i')$ from each constraint with true type $v_i$, and
a contribution $\alpha_i(v_i',v_i)$ from each constraint with report
$v_i$. Cancellation therefore gives the following flow conservation
equations characterizing \lemref[useful duals]{lem:useful};
the seller calculation is analogous.

\begin{definition}[Useful duals]\label[definition]{def:useful}
Let $\tau\ge0$ and let $\alpha_i(v_i,v_i'),\beta_j(c_j,c_j')\ge0$ for $v_i\in T_i$, $v_i'\in T_i^+$, $c_j\in T_j$ and $c_j'\in T_j^+$. Write
\[
  \Out_i(v_i):=\sum_{v_i'\in T_i^+}\alpha_i(v_i,v_i'),\qquad \In_i(v_i):=\sum_{v_i'\in T_i}\alpha_i(v_i',v_i),
\]
and define $\Out_j(c_j)$ and $\In_j(c_j)$ from $\beta_j$ in the same way. The triple $(\tau,\alpha,\beta)$ is a \emph{useful dual} if
\begin{align}
  \Out_i(v_i)&=\tau f_i(v_i)+\In_i(v_i)&&(i\in I,\ v_i\in T_i),\label{eq:buyer-flow}\\
  \Out_j(c_j)&=\tau g_j(c_j)+\In_j(c_j)&&(j\in J,\ c_j\in T_j).\label{eq:seller-flow}
\end{align}
\end{definition}

Read $\alpha_i$ as a flow on buyer $i$'s types: a source supplies $\tau f_i(v_i)$ to each type $v_i$, the edge $v_i\to v_i'$ carries $\alpha_i(v_i,v_i')$, and the null report $\bot$ is the sink. \Cref{eq:buyer-flow} is then flow conservation. A useful dual is thus a budget multiplier $\tau$ together with one flow per agent, whose supply at every type is $\tau$ times the prior probability of that type.

\paratitle{Virtual values} Once the transfers are gone, what remains of the \lemref[Lagrangian at a useful dual]{lem:lagrangian} is the expected gains from trade, rescaled by $1+\tau$, with every value and cost replaced by a virtual one.

\begin{definition}[Virtual values and virtual surplus]\label[definition]{def:transformed}
Let $\tau\ge0$ and let $\alpha$ and $\beta$ be nonnegative multipliers. For a buyer $i$ and a trade $e\in E_i$, and for a seller $j$ and a trade $e\in E_j$, the \emph{virtual value} and the \emph{virtual cost} are
\begin{equation}\label{eq:transformed}
  \begin{aligned}
  \tv_{i,e}(v_i,\alpha,\tau)
    &:=v_{i,e}
      -\frac{1}{(1+\tau)f_i(v_i)}
       \sum_{v_i'\in T_i}\alpha_i(v_i',v_i)
       \,(v'_{i,e}-v_{i,e}),\\
  \tc_{j,e}(c_j,\beta,\tau)
    &:=c_{j,e}
      -\frac{1}{(1+\tau)g_j(c_j)}
       \sum_{c_j'\in T_j}\beta_j(c_j',c_j)
       \,(c'_{j,e}-c_{j,e}),
  \end{aligned}
\end{equation}
and the \emph{virtual surplus} at a profile $(v,c)$ is
\[
  \tH(v,c,\alpha,\beta,\tau):=\max_{X\in\cF}\sum_{e\in X}\big(\tv_{i(e),e}(v_{i(e)},\alpha,\tau)-\tc_{j(e),e}(c_{j(e)},\beta,\tau)\big)\ge 0,
\]
the surplus $H$ of \Cref{eq:surplus} at virtual values; it is nonnegative because the empty set is feasible.
\end{definition}

We suppress the dual arguments when they are clear. The correction
in \Cref{eq:transformed} is an incentive adjustment:
$\alpha_i(v_i',v_i)$ weights the constraint against type $v_i'$
reporting $v_i$, and its contribution depends on the difference
$v'_{i,e}-v_{i,e}$. The adjustment to a coordinate may have either sign.

For a concrete example of a useful flow, consider the buyer in
bilateral trade. Let each value send all its flow to the next lower
support point $v^-$, with the lowest value sending its flow to $\bot$.
\Cref{eq:buyer-flow} gives
$\alpha(v,v^-)=\tau(1-F(v^-))$ for $v>v_{\min}$ and
$\alpha(v_{\min},\bot)=\tau$, where $F$ is the cumulative distribution
function. For every nonmaximal support point $v$,
\Cref{eq:transformed} therefore becomes
\[
  \tv(v)
    =v-\frac{\tau}{1+\tau}\cdot
       \frac{1-F(v)}{f(v)}\,(v^+-v),
\]
where $v^+$ is the next higher support point and $f(v)$ is the
probability mass at $v$. At the largest value,
$\tv(v_{\max})=v_{\max}$. The analogous seller flow moves to the next
higher cost, with the highest cost sending its flow to $\bot$. These are the canonical flows used by
\citet*{BrustleCaiWuZhao2017} to derive dual upper bounds
on gains from trade in bilateral trade.

\begin{theorem}[Duality]\label[theorem]{prop:dual}
At every useful dual $(\tau,\alpha,\beta)$,
\begin{equation}\label{eq:weak-duality}
  \SB\le (1+\tau)\,\E_{v,c}\big[\tH(v,c,\alpha,\beta,\tau)\big],
\end{equation}
and the second best is the least of these bounds,
\begin{equation}\label{eq:sb-min}
  \SB=\min_{(\tau,\alpha,\beta)\ \text{useful}}\ (1+\tau)\,\E_{v,c}\big[\tH(v,c,\alpha,\beta,\tau)\big],
\end{equation}
the minimum being attained.
\end{theorem}
The inequality is weak duality and the equality is strong
 duality; both are proved in \Cref{app:dual}.
It therefore suffices to show
\[
  \FB\le 2D,
  \qquad
  D:=(1+\tau)\E_{v,c}[\tH(v,c)],
\]
for every useful dual. \Cref{sec:virtual-map} uses the nonnegative
flow structure to represent the inverse virtual-value maps by
Markov chains, and flow conservation to control their terminal
distributions.

\subsection{The Markov recovery process}
\label{sec:virtual-map}

Fix a useful dual $(\tau,\alpha,\beta)$. Let $V_i$ and
$\widetilde V_i$ have rows indexed by $v_i\in T_i$ and columns
indexed by $e\in E_i$, with entries
\[
  V_i(v_i,e):=v_{i,e},
  \qquad
  \widetilde V_i(v_i,e):=\tv_{i,e}(v_i,\alpha,\tau).
\]
Define $C_j$ and $\widetilde C_j$ analogously for sellers.
The rows of these matrices are entire types or virtual-type vectors;
write
\[
  \tv_i(v_i,\alpha,\tau)
    :=\bigl(\tv_{i,e}(v_i,\alpha,\tau)\bigr)_{e\in E_i},
  \qquad
  \tc_j(c_j,\beta,\tau)
    :=\bigl(\tc_{j,e}(c_j,\beta,\tau)\bigr)_{e\in E_j}.
\]
A matrix acting from the left takes the same linear combination
of types in every trade coordinate.

We relate $\widetilde V_i$ to $V_i$, and $\widetilde C_j$ to $C_j$, through the matrices
\begin{equation}\label{eq:reverse-rates}
  \begin{aligned}
  \cA_i(v_i,v_i')&:=\frac{\alpha_i(v_i',v_i)}{(1+\tau)f_i(v_i)}\quad(v_i'\ne v_i),&\qquad \cA_i(v_i,v_i)&:=-\sum_{v_i'\ne v_i}\cA_i(v_i,v_i'),\\
  \cB_j(c_j,c_j')&:=\frac{\beta_j(c_j',c_j)}{(1+\tau)g_j(c_j)}\quad(c_j'\ne c_j),&\qquad \cB_j(c_j,c_j)&:=-\sum_{c_j'\ne c_j}\cB_j(c_j,c_j').
  \end{aligned}
\end{equation}

Writing $\Id$ for the identity matrix and substituting \Cref{eq:reverse-rates}, the row of $(\Id-\cA_i)V_i$ at $v_i$ is
\begin{align*}
  &v_i-\cA_i(v_i,v_i)\,v_i-\sum_{v_i'\ne v_i}\cA_i(v_i,v_i')\,v_i' =v_i-\frac{1}{(1+\tau)f_i(v_i)}\sum_{v_i'\in T_i}\alpha_i(v_i',v_i)\,(v_i'-v_i) =\tv_i(v_i,\alpha,\tau),
\end{align*}
the last equality being \Cref{eq:transformed}. On the seller side the computation is the same, so
\begin{equation}\label{eq:transformed-Q}
  \widetilde V_i=(\Id-\cA_i)V_i\quad(i\in I),\qquad \widetilde C_j=(\Id-\cB_j)C_j\quad(j\in J).
\end{equation}

To bound the first best by the second best, we reconstruct the true values and costs from the virtual ones, which means inverting $\Id-\cA_i$ and $\Id-\cB_j$. The inverse has an interpretable form: it is the transition matrix of a Markov chain on the agent's types, run for a random time. Each true value is therefore an average of virtual values, taken at the random state that the chain reaches from it. More precisely:

\begin{lemma}[Markov-chain recovery]\label[lemma]{lem:markov}
For every buyer $i$, the matrix $\Id-\cA_i$ is invertible, and
\begin{equation}\label{eq:resolvent}
  (\Id-\cA_i)^{-1}=\int_0^\infty e^{-\theta}e^{\theta\cA_i}\,\mathrm{d}\theta .
\end{equation}
The matrices $e^{\theta\cA_i}$, $\theta\ge0$, are the transition matrices of a continuous-time Markov chain on $T_i$: for every $\theta\ge0$, the row $e^{\theta\cA_i}(v_i,\cdot)$ is a probability distribution on $T_i$, the law of the state of the chain at time $\theta$ when it starts at $v_i$. By \Cref{eq:resolvent}, if $\theta$ is drawn from $\Exp(1)$, independently of the chain, the state $\hat v_i$ at time $\theta$ has law $(\Id-\cA_i)^{-1}(v_i,\cdot)$. Consequently, the virtual value at this state equals the true value on average:
\begin{equation}\label{eq:recovery}
  \mathop{\E}\limits_{\substack{\theta\sim\Exp(1)\\ \hat v_i\sim e^{\theta\cA_i}(v_i,\cdot)}}\big[\tv_i(\hat v_i,\alpha,\tau)\big]=v_i .
\end{equation}
The same holds for every seller $j$, with $\cB_j$, $c_j$ and $\tc_j$ in place of $\cA_i$, $v_i$ and $\tv_i$.
\end{lemma}

The proof is in \Cref{app:markov}.

The recovery identity evaluates virtual values at terminal types,
whereas the dual objective averages virtual surplus under the prior.
We therefore compare the terminal distribution with the prior.
Write $f_i$ and $g_j$ for the buyers' and sellers' prior mass functions,
viewed as row vectors. If buyer $i$'s initial type is drawn from $f_i$ independently of its
clock, then conditional on the clock taking the value $\theta$,
its terminal type has distribution
\[
    p_{i,\theta}:=f_i e^{\theta\cA_i}.
\]
Averaging over the $\Exp(1)$ clock gives the terminal law
$f_i(\Id-\cA_i)^{-1}$. We first establish a density bound for each
fixed $\theta$, retaining the time dependence needed to exploit
the coupling of clocks later.

Write $\kappa:=\tau/(1+\tau)$. The flow conservation constraints
bound the initial rate of change of the prior. Indeed, by
\Cref{eq:reverse-rates,eq:buyer-flow}, for every $v_i\in T_i$,
\begin{align*}
  (1+\tau)(f_i\cA_i)(v_i)
  &=\sum_{v_i'\ne v_i}\alpha_i(v_i,v_i')
    -\sum_{v_i'\ne v_i}\alpha_i(v_i',v_i)\\
  &=\Out_i(v_i)-\In_i(v_i)-\alpha_i(v_i,\bot)\\
  &=\tau f_i(v_i)-\alpha_i(v_i,\bot)
  \le \tau f_i(v_i).
\end{align*}
The same calculation applies to sellers, so every useful dual satisfies
\begin{equation}\label{eq:cap}
    f_i\cA_i\le\kappa f_i\quad(i\in I),
    \qquad
    g_j\cB_j\le\kappa g_j\quad(j\in J),
\end{equation}
where inequalities between row vectors hold componentwise.

Since $e^{\theta\cA_i}$ has nonnegative entries, this initial bound
controls the rate of change at every later time:
\[
    \frac{\mathrm{d}}{\mathrm{d}\theta}p_{i,\theta}
    =(f_i\cA_i)e^{\theta\cA_i}
    \le \kappa f_i e^{\theta\cA_i}
    =\kappa p_{i,\theta}.
\]
Thus the probability of each type grows at a rate of at most
$\kappa$ times its current probability. Starting from
$p_{i,0}=f_i$, this bounds its probability at time $\theta$ by
$e^{\kappa\theta}$ times its prior probability.

\begin{lemma}[Prior-density bounds]\label[lemma]{lem:density}
At every useful dual and for every $\theta\ge0$,
\begin{equation}\label{eq:density}
    f_i e^{\theta\cA_i}\le e^{\kappa\theta}f_i\quad(i\in I),
    \qquad
    g_j e^{\theta\cB_j}\le e^{\kappa\theta}g_j\quad(j\in J),
\end{equation}
where $\kappa=\tau/(1+\tau)$.
In particular, averaging over the exponential clock gives
\[
    f_i(\Id-\cA_i)^{-1}\le (1+\tau)f_i \quad(i\in I),
    \qquad
    g_j(\Id-\cB_j)^{-1}\le (1+\tau)g_j \quad(j\in J).
\]
\end{lemma}

The proof is in \Cref{app:markov}.

\section{Bilateral trade}
\label{sec:clocks}
We first illustrate the argument in the bilateral-trade setting, with one buyer, one seller, and one
possible trade. The buyer's value $v$ and seller's cost $c$ are
drawn independently from priors with mass functions $f$ and $g$
on $T_I$ and $T_J$. Thus,
\[
  H(v,c)=(v-c)^+,
  \qquad
  \FB=\E_{v,c}\big[(v-c)^+\big],
\]
where $w^+:=\max\{w,0\}$.

Fix an arbitrary useful dual $(\tau,\alpha,\beta)$.
Write $\tv$ and $\tc$ for its virtual value and virtual cost,
and let $\cA$ and $\cB$ be the corresponding Markov generators
from \Cref{sec:virtual-map}. The virtual surplus and dual
objective are
\[
  \tH(v,c)=\big(\tv(v)-\tc(c)\big)^+,
  \qquad
  D:=(1+\tau)\E_{v,c}\big[\tH(v,c)\big].
\]
By \thmref[duality]{prop:dual}, it suffices to prove
\begin{equation}\label{eq:bilateral-target}
  D\ge\tfrac12\,\FB.
\end{equation}

Choose any joint law $\mu$ for clocks $(\theta_B,\theta_S)$
with $\Exp(1)$ marginals, independently of the initial types
$(v,c)$. We will specify $\mu$ below.
Conditional on the initial types and the clocks, draw the
terminal types independently according to
\[
  \hat v\sim e^{\theta_B\cA}(v,\cdot),
  \qquad
  \hat c\sim e^{\theta_S\cB}(c,\cdot).
\]
By \lemref{lem:markov}, the exponential marginals give
\[
  \E[\tv(\hat v)\mid v,c]=v,
  \qquad
  \E[\tc(\hat c)\mid v,c]=c.
\]
Since $w\mapsto w^+$ is convex, Jensen's inequality yields
\[
  (v-c)^+
  =\big(\E[\tv(\hat v)-\tc(\hat c)\mid v,c]\big)^+
  \le\E[\tH(\hat v,\hat c)\mid v,c].
\]
Averaging over the initial types gives
\begin{equation}\label{eq:bilateral-step1}
  \FB\le\E[\tH(\hat v,\hat c)].
\end{equation}

It remains to compare the expected virtual surplus at the
terminal types with its expectation under the prior,
which equals $D/(1+\tau)$.
Write $\kappa:=\tau/(1+\tau)$.
Conditional on $(\theta_B,\theta_S)=(s,t)$, the terminal types
are independent: the initial types are independent, and the
two chains use independent randomness. Their laws satisfy
\[
  f_s:=fe^{s\cA}\le e^{\kappa s}f,
  \qquad
  g_t:=ge^{t\cB}\le e^{\kappa t}g
\]
by the \lemref[prior-density bounds]{lem:density}. Since $\tH\ge0$, it follows that
\begin{equation}\label{eq:bilateral-step2}
  \begin{aligned}
  \E[\tH(\hat v,\hat c)\mid\theta_B=s,\theta_S=t]
  &=\sum_{\hat v\in T_I,\ \hat c\in T_J}
    f_s(\hat v)g_t(\hat c)\,\tH(\hat v,\hat c)\\
  &\le e^{\kappa(s+t)}\E_{v,c}[\tH(v,c)].
  \end{aligned}
\end{equation}
Averaging over the clocks and combining with
\Cref{eq:bilateral-step1} gives
\[
  \FB
  \le\frac{\E_\mu[e^{\kappa(\theta_B+\theta_S)}]}{1+\tau}\,D.
\]
This holds for every joint clock law $\mu$ with $\Exp(1)$
marginals. To prove \Cref{eq:bilateral-target}, it therefore
suffices to choose such a law satisfying
\begin{equation}\label{eq:price}
  \E_\mu[e^{\kappa(\theta_B+\theta_S)}]\le2(1+\tau).
\end{equation}

Independent clocks are not enough: for them the expectation in \Cref{eq:price} factors as
\[
  \E_{\theta_B}\big[e^{\kappa\theta_B}\big]\,\E_{\theta_S}\big[e^{\kappa\theta_S}\big]=(1+\tau)^2,
\]
which exceeds $2(1+\tau)$ whenever $\tau>1$.

Instead, draw a single $u\sim\Unif(0,1)$, independently of the
initial types and chain randomness, and couple the clocks
antithetically:
\begin{equation}\label{eq:antithetic}
  \theta_B(u):=-\log u,
  \qquad
  \theta_S(u):=-\log(1-u).
\end{equation}
Both marginals are $\Exp(1)$: for $\theta\ge0$,
\[
  \Pr[\theta_B(u)>\theta]
  =\Pr[u<e^{-\theta}]
  =e^{-\theta},
\]
and similarly for $\theta_S(u)$. A long buyer clock is paired
with a short seller clock, and the expectation in
\Cref{eq:price} becomes
\[
  \E_u\big[e^{\kappa(\theta_B(u)+\theta_S(u))}\big]
  =\int_0^1u^{-\kappa}(1-u)^{-\kappa}\,\mathrm{d}u.
\]
This is Euler's beta integral
$\mathrm{B}(1-\kappa,1-\kappa)$, where
\[
  \mathrm{B}(a,b):=\int_0^1u^{a-1}(1-u)^{b-1}\,\mathrm{d}u
  \qquad(a,b>0).
\]
The following lemma bounds it using
$1-\kappa=1/(1+\tau)$.
\begin{lemma}[Beta-integral bound]\label[lemma]{lem:beta}
For $\tau\ge0$ and $\kappa=\tau/(1+\tau)$,
\[
  \Beta(\tau):=\int_0^1u^{-\kappa}(1-u)^{-\kappa}\,\mathrm{d}u<2(1+\tau).
\]
\end{lemma}

\begin{proof}
Multiplying the integrand by $1=u+(1-u)$ and using $u^{1-\kappa}<1$ and $(1-u)^{1-\kappa}<1$ on $(0,1)$ gives
\[
  u^{-\kappa}(1-u)^{-\kappa}<(1-u)^{-\kappa}+u^{-\kappa}\qquad(u\in(0,1)),
\]
and each term on the right integrates to $\int_0^1u^{-\kappa}\,\mathrm{d}u=\frac{1}{1-\kappa}=1+\tau$.
\end{proof}

By the \lemref[beta-integral bound]{lem:beta}, the clocks defined by \Cref{eq:antithetic} satisfy \Cref{eq:price}. This proves \Cref{eq:bilateral-target}, that is, $\SB\ge\frac12\FB$ for bilateral trade.

\section{General case: multidimensional markets}
\label{sec:market}
We now extend the bilateral-trade argument of \Cref{sec:clocks}
to general markets. The additional ingredient is a distributional
comparison that loses only one density factor per side,
regardless of the number of agents.

Fix a useful dual $(\tau,\alpha,\beta)$ and write
\[
  D:=(1+\tau)\E_{v,c}[\tH(v,c)],
  \qquad
  \kappa:=\frac{\tau}{1+\tau}.
\]
By \thmref[duality]{prop:dual}, it suffices to prove
\begin{equation}\label{eq:market-target}
  D\ge\tfrac12\,\FB.
\end{equation}
Let $\cA_i$ and $\cB_j$ be the corresponding Markov generators
from \Cref{sec:virtual-map}.

Choose a joint law $\mu$ for clocks $(\theta_B,\theta_S)$
with $\Exp(1)$ marginals, independently of the initial type
profile $(v,c)$. All buyers share $\theta_B$, and all sellers
share $\theta_S$. Conditional on the initial types and the
clocks, draw the terminal types independently across agents:
\[
  \hat v_i\sim e^{\theta_B\cA_i}(v_i,\cdot)
  \quad(i\in I),
  \qquad
  \hat c_j\sim e^{\theta_S\cB_j}(c_j,\cdot)
  \quad(j\in J).
\]
By \lemref{lem:markov},
\[
  \E[\tv_i(\hat v_i)\mid v,c]=v_i
  \quad(i\in I),
  \qquad
  \E[\tc_j(\hat c_j)\mid v,c]=c_j
  \quad(j\in J).
\]
Thus the expected virtual gain of each trade at the terminal
profile equals its true gain. Since maximum feasible surplus
is a convex function of the vector of trade gains, Jensen's
inequality gives
\[
  H(v,c)\le\E[\tH(\hat v,\hat c)\mid v,c].
\]
Averaging over the initial types,
\begin{equation}\label{eq:market-step1}
  \FB\le\E[\tH(\hat v,\hat c)].
\end{equation}

As in bilateral trade, we compare expected virtual surplus at
the terminal profile with its expectation under the prior.
Conditional on the clocks, the terminal types are independent
across agents, with laws $f_ie^{\theta_B\cA_i}$ and
$g_je^{\theta_S\cB_j}$. By the \lemref[prior-density bounds]{lem:density},
\begin{equation}\label{eq:market-density}
  f_ie^{\theta_B\cA_i}\le e^{\kappa\theta_B}f_i
  \quad(i\in I),
  \qquad
  g_je^{\theta_S\cB_j}\le e^{\kappa\theta_S}g_j
  \quad(j\in J).
\end{equation}
Multiplying these bounds over all agents would incur the factor
$e^{\kappa(|I|\theta_B+|J|\theta_S)}$.
The following lemma uses downward closure to retain only
one density factor per side.

\begin{lemma}[One factor per side]\label[lemma]{lem:one-factor}
Conditional on the clocks,
\[
  \E[\tH(\hat v,\hat c)\mid\theta_B,\theta_S]
  \le e^{\kappa(\theta_B+\theta_S)}
      \E_{v,c}[\tH(v,c)].
\]
\end{lemma}

\begin{proof}
Fix the clocks and write $K_B:=e^{\kappa\theta_B}$ and
$K_S:=e^{\kappa\theta_S}$. All expectations in this proof
are conditional on these clock values.
We compare the terminal profile with an auxiliary profile
$(\bar v,\bar c)$ drawn from the original product prior,
whose expected virtual surplus is $D/(1+\tau)$.
We replace one side at a time.

\paratitle{Buyers}
Fix a sellers' profile $\hat c$ and write
$\hat f_i:=f_ie^{\theta_B\cA_i}$.
By \Cref{eq:market-density}, $\hat f_i\le K_Bf_i$.
We show that
\begin{equation}\label{eq:thinning}
  \E_{\hat v}[\tH(\hat v,\hat c)]
  \le K_B\,\E_{\bar v}[\tH(\bar v,\hat c)],
\end{equation}
where the $\hat v_i$ have independent laws $\hat f_i$ and
the auxiliary types $\bar v_i$ have independent laws $f_i$.

Only these marginal laws enter \Cref{eq:thinning}, so we may
couple the profiles as follows. Independently for each buyer
$i$, draw $\bar v_i$ from $f_i$ and a coin $\chi_i\in\{0,1\}$
with
\[
  \Pr[\chi_i=1\mid\bar v_i]
  =\frac{\hat f_i(\bar v_i)}{K_Bf_i(\bar v_i)}.
\]
This probability is at most $1$ because $\hat f_i\le K_Bf_i$.
If $\chi_i=1$, set $\hat v_i=\bar v_i$; otherwise, draw
$\hat v_i$ afresh from $\hat f_i$. For every $t\in T_i$,
\[
  \Pr[\chi_i=1,\hat v_i=t]=\frac{\hat f_i(t)}{K_B}.
\]
Hence $\Pr[\chi_i=1]=1/K_B$, and
\[
  \Pr[\hat v_i=t]
  =\frac{\hat f_i(t)}{K_B}
   +\left(1-\frac{1}{K_B}\right)\hat f_i(t)
  =\hat f_i(t).
\]
Thus $\chi_i$ is independent of $\hat v_i$, has mean $1/K_B$,
and guarantees $\bar v_i=\hat v_i$ whenever $\chi_i=1$.
The independent construction across buyers makes each coin
independent of the entire terminal buyers' profile $\hat v$.

Choose a feasible set $X^*=X^*(\hat v,\hat c)$ maximizing
virtual surplus at $(\hat v,\hat c)$, and retain
\[
  X':=\{e\in X^*:\chi_{i(e)}=1\}.
\]
Downward closure gives $X'\in\cF$. Every retained trade has
the same virtual gain at $(\bar v,\hat c)$ as at
$(\hat v,\hat c)$, so
\[
  \tH(\bar v,\hat c)
  \ge\sum_{e\in X^*}\chi_{i(e)}
    \big(\tv_{i(e),e}(\hat v_{i(e)})
       -\tc_{j(e),e}(\hat c_{j(e)})\big).
\]
Taking conditional expectations given $\hat v$ yields
\[
  \E[\tH(\bar v,\hat c)\mid\hat v]
  \ge\frac{1}{K_B}\tH(\hat v,\hat c).
\]
Averaging proves \Cref{eq:thinning}. Each trade is controlled
by one buyer's coin, so the loss is $K_B$ regardless of the
number of buyers.

\paratitle{Sellers}
For each seller $j$, the terminal law
$\hat g_j:=g_je^{\theta_S\cB_j}$ satisfies
$\hat g_j\le K_Sg_j$.
The same argument, for each fixed buyers' profile $\bar v$,
gives
\[
  \E_{\hat c}[\tH(\bar v,\hat c)]
  \le K_S\,\E_{\bar c}[\tH(\bar v,\bar c)],
\]
where the auxiliary seller types $\bar c_j$ have independent
laws $g_j$. Averaging the two comparisons over the other
side's profile gives
\[
  \E_{\hat v,\hat c}[\tH(\hat v,\hat c)]
  \le K_B\,\E_{\bar v,\hat c}[\tH(\bar v,\hat c)]
  \le K_BK_S\,\E_{\bar v,\bar c}[\tH(\bar v,\bar c)].
\]
Here each expectation uses the indicated product laws.
Since $(\bar v,\bar c)$ has the original product prior,
this proves the lemma.
\end{proof}

\begin{proof}[Proof of \thmref{thm:main}]
Choose the shared clocks according to the antithetic coupling
in \Cref{eq:antithetic}. Combining \Cref{eq:market-step1}
with the bound of \lemref[one factor per side]{lem:one-factor} gives
\[
  \FB
  \le \E_\mu[e^{\kappa(\theta_B+\theta_S)}]\,
      \E_{v,c}[\tH(v,c)]
  =\frac{\Beta(\tau)}{1+\tau}\,D
  \le 2D,
\]
where the last inequality follows from the \lemref[beta-integral bound]{lem:beta}.
This proves \Cref{eq:market-target} for every useful dual.
Taking the minimum over useful duals by \thmref[duality]{prop:dual}
yields $\SB\ge\tfrac12\FB$.
Finally, \propref[budget-balance equivalence]{prop:payment} ensures that the second best
is attained by a mechanism in $\cM$, completing the proof.
\end{proof}

\section{Conclusion}
\label{sec:conclusion}

We establish that the second best is always at least half of
the first best in multidimensional two-sided markets with
arbitrary within-agent correlation and downward-closed
feasibility constraints, and that this guarantee is tight.
Our proof bounds every feasible dual through a connection
to continuous-time Markov chains, avoiding a structural
characterization of an optimal mechanism or dual. A natural
direction for future work is to explore whether this connection
can also yield approximation guarantees in other multidimensional
mechanism-design problems.

\section*{Acknowledgments}
Dirk Bergemann acknowledges financial support from NSF SES 2519400 and ONR MURI N00014-24-1-2742. Yang Cai and Jinzhao Wu were supported by NSF Award CCF-2342642, and Jinzhao Wu additionally by a CADMY Research Fellowship. Renzhe Xu was supported by the National Natural Science Foundation of China under Grants 72442024 and 62602401. This work was carried out while Kangning Wang was visiting Shanghai University of Finance and Economics.

\section*{AI methodology and disclosure}
We began this project in early 2026 by seeking to generalize the work of \citet*{BergemannHeumannMorris2023, BergemannHeumannMorris2025}, \emph{Cost Based Nonlinear Pricing} and \emph{Procurement without Priors: A Simple Mechanism
and its Notable Performance}, and have held weekly research discussions since March 2026. Our guiding question has been whether, in certain generalized settings, the second best always achieves a constant fraction of the first best. Initially, we obtained results in specific settings and made partial progress in broader settings through human research. This phase continued until May 2026, when GPT-5.5 produced a counterexample to a claim we had hoped to prove. For some of us, this was our first personal experience of AI resolving a problem to which we had devoted serious effort. From that turning point onward, as AI systems became more capable, the technical derivations became a mixed effort between the authors and AI. In these interactions, the authors supplied the AI systems with research notes, including conjectures, results already obtained, and partial proof attempts.

On August 10, 2026, GPT-5.6 Sol Ultra resolved one of our conjectures in the affirmative after working for 30 hours in a single session. The conjecture concerned an intermediate setting with a single unit-demand buyer, a single unit-supply seller, and multiple possible items to trade; its formulation drew on ideas from the work of \citet*{BergemannHeumannMorris2023,BergemannHeumannMorris2025}. To our surprise, the model found a short proof with an approximation factor of $2$, which matches, even in the simplest bilateral-trade setting, the optimal factor in the then-recent work of \citet*{LiuQinRenWang2026}. We spent the following weeks digesting the ideas, connecting the results to the literature, exploring how broadly the argument extends, and reformulating and presenting AI-generated proofs in ways that we find more natural. During the writing phase, we also used Claude Opus and ChatGPT to help draft and edit parts of the paper.

\setlength{\bibsep}{3pt plus 1pt minus 1pt}
\bibliographystyle{bilateral-fullauthors}
\bibliography{ref}

\newpage
\appendix

\section{Omitted proofs}
\label[appendix]{app:proofs}

\subsection{The second-best program and payment normalization}
\label[appendix]{app:payment}

Fix a market $\cN$ with priors $(F_i)_{i\in I}$ and $(G_j)_{j\in J}$, over which every expectation below is taken. For the null report $\bot$ of \Cref{sec:program}, representing nonparticipation, fix
\[
  x_{E_i}(\bot,v_{-i},c):=0,\quad p_i(\bot,v_{-i},c):=0,\qquad x_{E_j}(v,\bot,c_{-j}):=0,\quad p_j(v,\bot,c_{-j}):=0,
\]
so that an agent who reports $\bot$ trades nothing and neither pays nor receives anything. With $T_i^+$ and $T_j^+$ as in \Cref{sec:program}, the second best is characterized by \Cref{prog:second-best}, the full form of \Cref{eq:lp-compact}.

\begin{programbox}
\refstepcounter{program}\label[program]{prog:second-best}
\statementheading{{\bfseries\hyperref[prog:second-best]{Program~(\theprogram)}}\enspace\textemdash\enspace Second-best program}

\setlength{\leftmargini}{1.4em}\textbf{Variables:}
\begin{itemize}
\item $x_e(v,c)$, for all trades $e\in E$ and all profiles $(v,c)\in T_I\times T_J$, denoting the probability that trade $e$ is executed when the reports are $(v,c)$.
\item $p_i(v,c)$ and $p_j(v,c)$, for all buyers $i\in I$, sellers $j\in J$ and profiles $(v,c)$, denoting buyer $i$'s payment and seller $j$'s receipt at that profile, of either sign.
\end{itemize}
\textbf{Constraints:}
\begin{itemize}
\item for all $i\in I$, $v_i\in T_i$ and $v_i'\in T_i^+$,
\begin{equation}\label{eq:aux-buyer}
  \E_{v_{-i},c}\Big[\big(v_i\cdot x_{E_i}(v_i,v_{-i},c)-p_i(v_i,v_{-i},c)\big)-\big(v_i\cdot x_{E_i}(v_i',v_{-i},c)-p_i(v_i',v_{-i},c)\big)\Big]\ge0,
\end{equation}
guaranteeing that the mechanism is BIC and interim IR for every buyer;
\item for all $j\in J$, $c_j\in T_j$ and $c_j'\in T_j^+$,
\begin{equation}\label{eq:aux-seller}
  \E_{v,c_{-j}}\Big[\big(p_j(v,c_j,c_{-j})-c_j\cdot x_{E_j}(v,c_j,c_{-j})\big)-\big(p_j(v,c_j',c_{-j})-c_j\cdot x_{E_j}(v,c_j',c_{-j})\big)\Big]\ge0,
\end{equation}
guaranteeing the same for every seller;
\item ex-ante weak budget balance,
\begin{equation}\label{eq:aux-budget}
  \E_{v,c}\Big[\sum_{i\in I}p_i(v,c)-\sum_{j\in J}p_j(v,c)\Big]\ge0;
\end{equation}
\item feasibility of the allocation,
\begin{equation}\label{eq:aux-feasible}
  x(v,c)\in\cX(\cF)\qquad\text{for all profiles }(v,c)\in T_I\times T_J.
\end{equation}
\end{itemize}
\textbf{Objective:}
\begin{itemize}
\item the expected gains from trade,
\[
  \max\ \ \GFT(x)=\E_{v,c}\Big[\sum_{e\in E}x_e(v,c)\big(v_{i(e),e}-c_{j(e),e}\big)\Big].
\]
\end{itemize}
\end{programbox}

The constraints of \Cref{eq:aux-buyer} and \Cref{eq:aux-seller} whose report is a type are \Cref{eq:bic}; those whose report is $\bot$ read $\cU_i(v_i;v_i)\ge0$ and $\cU_j(c_j;c_j)\ge0$, which is \Cref{eq:ir}. We call them the \emph{incentive} and the \emph{participation} constraints of \Cref{prog:second-best}.

Since $\cX(\cF)$ is the convex hull of finitely many points and every constraint is linear in the allocation rule and the payments, \Cref{prog:second-best} is a finite linear program.

\begin{proposition}[Budget-balance equivalence]\label[proposition]{prop:payment}
Let $\cM$ be the class of direct mechanisms satisfying \Cref{eq:bic}, \Cref{eq:ir} and \Cref{eq:sbb}, and let $\cM'$ be obtained from it by dropping the requirement that transfers be nonnegative and replacing strong budget balance in \Cref{eq:sbb} by ex-ante weak budget balance, formally:
\[
  \E_{v,c}\Big[\sum_{i\in I}p_i(v,c)\Big]\ge\E_{v,c}\Big[\sum_{j\in J}p_j(v,c)\Big].
\]
Then the two classes implement the same allocation rules: an allocation rule is the allocation rule of some mechanism in $\cM'$ if and only if it is the allocation rule of some mechanism in $\cM$. In particular
\[
  \SB=\sup_{M=(x,p)\in\cM}\GFT(x)=\sup_{M=(x,p)\in\cM'}\GFT(x),
\]
and both suprema are attained.
\end{proposition}

\begin{lemma}[Second best as a linear program]\label[lemma]{lem:sb-is-lp}
The optimum of \Cref{prog:second-best} is attained and equals the second best,
\[
  \SB=\mathrm{OPT}(\programsymbol),
\]
and every optimal allocation rule of \Cref{prog:second-best} is the allocation rule of a mechanism in $\cM$ with gains from trade $\SB$.
\end{lemma}

The budget-balance conversion follows \citet*[Lemma~11 and Theorem~6]{BrustleCaiWuZhao2017}, whose allocation-preserving transformation applies to general two-sided markets with nonnegative transfers. We give a proof for the formulation used here: first normalize signed transfers, then balance the expected totals and obtain reportwise strong budget balance through a common pool. The common-pool construction below is an alternative to their product construction.

Write (a) for the statement that an allocation rule $x:T_I\times T_J\to\cX(\cF)$ is the allocation rule of a mechanism in $\cM'$, and (b) for the statement that it is the allocation rule of a mechanism in $\cM$. \propref{prop:payment} is the equivalence of the two. Since the feasible points of \Cref{prog:second-best} are exactly the mechanisms of $\cM'$, it is also what identifies $\SB$ with the value of \Cref{prog:second-best} when writing the \lemref[second best as a linear program]{lem:sb-is-lp}. Three lemmas isolate the moves of the proof: transfers may be taken to depend on the agent's own report alone, they may be shifted by constants, and nonnegative transfers may be routed through a common pool. A fourth records that \Cref{prog:second-best} is a finite linear program.

\begin{lemma}[Own-report transfers]\label[lemma]{lem:own}
For an allocation rule $x$ and real-valued transfers $p$ put
\[
  \hat p_i(v,c):=\E_{v_{-i},c}\big[p_i(v_i,v_{-i},c)\big],\qquad \hat p_j(v,c):=\E_{v,c_{-j}}\big[p_j(v,c_j,c_{-j})\big],
\]
which depend on $v_i$ and on $c_j$ alone. Then $\hat M:=(x,\hat p)$ has the same interim utilities as $M$, and $\E_{v,c}[\hat p_i(v,c)]=\E_{v,c}[p_i(v,c)]$ and $\E_{v,c}[\hat p_j(v,c)]=\E_{v,c}[p_j(v,c)]$. In particular $\hat M$ is BIC and interim IR whenever $M$ is, and satisfies ex-ante weak budget balance whenever $M$ does.
\end{lemma}

\begin{proof}
At true type $v_i$ and report $v_i'$, buyer $i$'s interim utility depends on the transfer only through $\E_{v_{-i},c}[p_i(v_i',v_{-i},c)]$, which is the constant value of $\hat p_i$ at that report; so the two mechanisms have the same $\cU_i$, and likewise the same $\cU_j$. For the expectations, independence across agents makes $\prod_{i'\ne i}f_{i'}(v_{i'})\prod_{j\in J}g_j(c_j)$ the joint mass function of the other agents' types, so
\[
  \E_{v,c}\big[p_i(v,c)\big] =\sum_{v_i\in T_i}f_i(v_i)\sum_{v_{-i},c}\prod_{i'\ne i}f_{i'}(v_{i'})\prod_{j\in J}g_j(c_j)\,p_i(v_i,v_{-i},c) =\E_{v,c}\big[\hat p_i(v,c)\big],
\]
and likewise for sellers. The three requirements defining $\cM'$ depend on the transfers only through these two objects.
\end{proof}

Using \lemref[own-report transfers]{lem:own}, we assume, wherever it is convenient below, that $p_i(v,c)$ depends on $v_i$ alone and $p_j(v,c)$ on $c_j$ alone; we then write $p_i(v_i)$ and $p_j(c_j)$.

\begin{lemma}[Constant shifts]\label[lemma]{lem:shift}
Let $M=(x,p)$ be BIC and interim IR and let $\gamma_i,\delta_j\in\R$. Replacing $p_i$ by $p_i+\gamma_i$ and $p_j$ by $p_j-\delta_j$ lowers every interim utility $\cU_i(v_i;v_i')$ by $\gamma_i$ and every $\cU_j(c_j;c_j')$ by $\delta_j$, so the mechanism stays BIC; it stays interim IR if and only if $\gamma_i\le m_i:=\min_{v_i\in T_i}\cU_i(v_i;v_i)$ for every $i$ and $\delta_j\le n_j:=\min_{c_j\in T_j}\cU_j(c_j;c_j)$ for every $j$.
\end{lemma}

\begin{proof}
A constant added to buyer $i$'s transfer is subtracted from each of its interim utilities, whatever the report, and a constant subtracted from seller $j$'s receipt is subtracted from each of its interim utilities. An incentive constraint compares two such utilities of the same agent, so it is unaffected; participation compares one of them with zero, and the minimal truthful utilities become $m_i-\gamma_i$ and $n_j-\delta_j$.
\end{proof}

\begin{lemma}[Common-pool transfers]\label[lemma]{lem:pool}
Let $p_i:T_i\to\R_+$ and $p_j:T_j\to\R_+$ satisfy $\sum_i\E_{v_i}[p_i(v_i)]=\sum_j\E_{c_j}[p_j(c_j)]=:N$. There are finite nonnegative transfers $\tilde p$ satisfying reportwise SBB with $\E_{v_{-i},c}[\tilde p_i(v_i',v_{-i},c)]=p_i(v_i')$ for all $i$ and $v_i'$, and $\E_{v,c_{-j}}[\tilde p_j(v,c_j',c_{-j})]=p_j(c_j')$ for all $j$ and $c_j'$. If $N>0$ one may take $\tilde p_i(v,c)=p_i(v_i)S(c)/N$ and $\tilde p_j(v,c)=p_j(c_j)B(v)/N$ with $B(v):=\sum_ip_i(v_i)$ and $S(c):=\sum_jp_j(c_j)$; if $N=0$ all transfers vanish and zero transfers work.
\end{lemma}

\begin{proof}
If $N=0$, then $\sum_i\sum_{v_i}f_i(v_i)p_i(v_i)=0$ with $f_i>0$ and $p_i\ge0$ forces $p_i\equiv0$ for every $i$, and likewise $p_j\equiv0$; zero transfers have zero interim expectations and are balanced. If $N>0$, the displayed transfers are finite and nonnegative as products and quotients of finite nonnegative numbers, and at every profile
\[
  \sum_i\tilde p_i(v,c)=\frac{S(c)}{N}\sum_ip_i(v_i)=\frac{B(v)S(c)}{N}=\frac{B(v)}{N}\sum_jp_j(c_j)=\sum_j\tilde p_j(v,c),
\]
which is \Cref{eq:sbb}. After the report $v_i'$ of buyer $i$, the payment $\tilde p_i(v_i',v_{-i},c)=p_i(v_i')S(c)/N$ depends on $c$ alone, and $\E_c[S(c)]=\sum_j\E_{c_j}[p_j(c_j)]=N$ by the hypothesis, so $\E_{v_{-i},c}[\tilde p_i(v_i',v_{-i},c)]=p_i(v_i')$. Symmetrically $\E_v[B(v)]=N$ gives $\E_{v,c_{-j}}[\tilde p_j(v,c_j',c_{-j})]=p_j(c_j')$.
\end{proof}

\begin{lemma}[Finite linear program]\label[lemma]{lem:auxlp}
In the variables $x_e(v,c)$, $p_i(v,c)$ and $p_j(v,c)$, for $e\in E$, $i\in I$, $j\in J$ and $(v,c)\in T_I\times T_J$, \Cref{prog:second-best} is a linear program with finitely many variables and constraints. It is feasible ($x\equiv0$, $p\equiv0$), its objective is bounded above by $\FB$, and hence it has an optimal solution.
\end{lemma}

\begin{proof}
The interim utilities are linear in $(x,p)$ and the types that multiply the allocation in \Crefrange{eq:aux-buyer}{eq:aux-seller} are data, so the objective and every constraint is linear; there are $\big(|E|+|I|+|J|\big)\cdot|T_I\times T_J|$ variables. The feasibility requirement in \Cref{eq:aux-feasible} asks that $x(v,c)$ lie in $\cX(\cF)=\conv\{\1_X:X\in\cF\}$, the convex hull of finitely many points; such a set is a polytope and hence, by the Minkowski--Weyl theorem, the solution set of finitely many linear inequalities, so there are finitely many constraints as well. With $x\equiv0$ and zero transfers, \Crefrange{eq:aux-buyer}{eq:aux-budget} all read $0\ge0$, and \Cref{eq:aux-feasible} holds because $\varnothing\in\cF$ by downward closure and $\1_\varnothing=0$. The objective is bounded by $\FB$: $x(v,c)$ is a convex combination of the vectors $\1_X$, so $\sum_{e\in E}x_e(v,c)\big(v_{i(e),e}-c_{j(e),e}\big)\le H(v,c)$ at every profile, so $\GFT(x)\le\E_{v,c}[H(v,c)]=\FB$ for every allocation rule. A feasible linear program whose objective is bounded above has an optimal solution; that the feasible region is unbounded in the directions of the transfers is irrelevant.
\end{proof}

\begin{proof}[Proof of \propref{prop:payment} and \lemref{lem:sb-is-lp}]\leavevmode\par
We consider the two directions in turn.

\paratitle{(b)$\Rightarrow$(a)} Finite nonnegative transfers with reportwise SBB satisfy
\[
  \E_{v,c}\big[\sum_ip_i(v,c)\big]=\E_{v,c}\big[\sum_jp_j(v,c)\big],
\]
so a mechanism of $\cM$ lies in $\cM'$.

\paratitle{(a)$\Rightarrow$(b)} Let $M=(x,p)$ be BIC and interim IR, with real-valued transfers satisfying ex-ante weak budget balance. If $E=\varnothing$, then $\cF=\{\varnothing\}$ and $\cX(\cF)=\{0\}$, so $x\equiv0$; every support is a one-point set, and the mechanism with zero transfers is BIC (there are no deviations), interim IR and reportwise SBB, with allocation $x$. Assume $E\ne\varnothing$; then $I\ne\varnothing$ and $J\ne\varnothing$. Using \lemref[own-report transfers]{lem:own}, we may assume $p_i(v,c)=p_i(v_i)$ and $p_j(v,c)=p_j(c_j)$.

\paratitle{Normalization} Let $m_i$ and $n_j$ be defined as in \lemref[constant shifts]{lem:shift}; they are nonnegative because $M$ is interim IR. Applying \lemref[constant shifts]{lem:shift} with $\gamma_i=m_i$ and $\delta_j=n_j$: the transfers $p_i+m_i$ and $p_j-n_j$ leave the allocation and the incentive constraints untouched, and the minimal truthful utility of every agent is now zero. They are also nonnegative. For buyer $i$, choose $v_i^\circ$ attaining $m_i$ before the shift; the corresponding incentive constraint before the shift, at true type $v_i^\circ$ and report $v_i'$, reads $m_i=\cU_i(v_i^\circ;v_i^\circ)\ge\cU_i(v_i^\circ;v_i')$, which after adding $p_i(v_i')$ to both sides gives
\[
  p_i(v_i')+m_i\ge v_i^\circ\cdot\E_{v_{-i},c}\big[x_{E_i}(v_i',v_{-i},c)\big]\ge 0\qquad(v_i'\in T_i),
\]
the last inequality because $v_i^\circ\in\R_+^{E_i}$ and the expected allocation lies in $[0,1]^{E_i}$, being an average of vectors in $\cX(\cF)\subseteq[0,1]^E$; this is the only use of nonnegative values in the paper. (For $v_i'=v_i^\circ$ the incentive step is vacuous and the inequality holds directly; if $|T_i|=1$ there is no nontrivial incentive constraint and the same formula applies; if $E_i=\varnothing$ the support is a one-point set and $p_i+m_i\equiv0$.) For seller $j$,
\[
  p_j(c_j)-n_j=c_j\cdot\E_{v,c_{-j}}\big[x_{E_j}(v,c_j,c_{-j})\big]+\big(\cU_j(c_j;c_j)-n_j\big)\ge 0\qquad(c_j\in T_j),
\]
because the nonnegativity of $c_j$ and of the expected allocation makes the first term nonnegative, and the definition of $n_j$ makes the second nonnegative; here no incentive constraint is used, only that definition and $c_j\ge0$.

\paratitle{Balancing the expected totals} The shifts raise expected buyer payments and lower expected seller receipts, so, writing $p$ again for the shifted transfers,
\[
  \Delta:=\sum_i\E_{v_i}[p_i(v_i)]-\sum_j\E_{c_j}[p_j(c_j)]\ge 0
\]
by ex-ante weak budget balance, the shifts having increased $\Delta$ by $\sum_im_i+\sum_jn_j\ge0$. Choose one seller $j_0$ (possible since $J\ne\varnothing$) and replace $p_{j_0}$ by $p_{j_0}+\Delta$; applying \lemref[constant shifts]{lem:shift} with $\delta_{j_0}=-\Delta\le0$, this preserves BIC and interim IR, and it preserves nonnegativity. Denoting the resulting transfers again by $p$,
\[
  \sum_i\E_{v_i}[p_i(v_i)]=\sum_j\E_{c_j}[p_j(c_j)]=:N\ge 0 .
\]
The allocation has not changed in any of these steps.

\paratitle{Pooling} \lemref{lem:pool}, applied to the nonnegative $p$ with the common total $N$, yields finite nonnegative transfers $\tilde p$ satisfying reportwise SBB whose own-report averages are again $p$. By \lemref[own-report transfers]{lem:own}, the mechanism $(x,\tilde p)$ therefore has the same interim utilities as $(x,p)$, hence is BIC and interim IR. Its allocation is $x$, which proves (b).

\paratitle{Value and attainment} The feasible points of \Cref{prog:second-best} are exactly the mechanisms of $\cM'$, so $\mathrm{OPT}(\programsymbol)=\sup_{M=(x,p)\in\cM'}\GFT(x)$, and this supremum is attained because \Cref{prog:second-best} is a \lemref[finite linear program]{lem:auxlp}. By (b)$\Rightarrow$(a) every mechanism of $\cM$ lies in $\cM'$, so $\SB\le\mathrm{OPT}(\programsymbol)$; conversely (a)$\Rightarrow$(b), applied to an optimal point of \Cref{prog:second-best}, produces a mechanism of $\cM$ with the same allocation rule and hence the same gains from trade, so $\SB\ge\mathrm{OPT}(\programsymbol)$ and the supremum over $\cM$ is attained as well. This proves the displayed equalities for \propref[budget-balance equivalence]{prop:payment} and the \lemref[second best as a linear program]{lem:sb-is-lp}. If $\cF=\{\varnothing\}$ with $E\ne\varnothing$, all of this holds with $x\equiv0$ and $\FB=\SB=0$.
\end{proof}

\subsection{Duality framework}
\label[appendix]{app:dual}

We write out the partial Lagrangian behind \Cref{sec:program}, show that its transfers cancel exactly at the useful duals of \Cref{def:useful}, record its coefficients, and prove \thmref[duality]{prop:dual}. Throughout, $\Out_i(v_i)$, $\In_i(v_i)$, $\Out_j(c_j)$ and $\In_j(c_j)$ are the multiplier sums of \Cref{def:useful}; the multiplier vector is $z:=(\tau,\alpha,\beta)$; and $\pi(v,c):=\prod_{i\in I}f_i(v_i)\prod_{j\in J}g_j(c_j)$ is the probability of the profile $(v,c)$.

We dualize \Cref{prog:second-best} as in \citet*{BrustleCaiWuZhao2017}. Attach a nonnegative multiplier to each constraint except \Cref{eq:aux-feasible}, which we keep as the domain: $\tau$ to the budget constraint in \Cref{eq:aux-budget}, $\alpha_i(v_i,v_i')$ to the constraint in \Cref{eq:aux-buyer} with true type $v_i$ and report $v_i'$, and $\beta_j(c_j,c_j')$ to its seller counterpart in \Cref{eq:aux-seller}. The partial Lagrangian is
\begin{equation}\label{eq:lagrangian}
  \begin{aligned}
  \cL(x,p;\tau,\alpha,\beta) &=\GFT(x)+\tau\,\E_{v,c}\Big[\sum_{i\in I}p_i(v,c)-\sum_{j\in J}p_j(v,c)\Big]\\
  &\quad+\sum_i\ \sum_{v_i\in T_i,\,v_i'\in T_i^+}\alpha_i(v_i,v_i')\,\E_{v_{-i},c}\Big[\big(v_i\cdot x_{E_i}(v_i,v_{-i},c)-p_i(v_i,v_{-i},c)\big)\\
  &\hspace{12em}-\big(v_i\cdot x_{E_i}(v_i',v_{-i},c)-p_i(v_i',v_{-i},c)\big)\Big]\\
  &\quad+\sum_j\ \sum_{c_j\in T_j,\,c_j'\in T_j^+}\beta_j(c_j,c_j')\,\E_{v,c_{-j}}\Big[\big(p_j(v,c_j,c_{-j})-c_j\cdot x_{E_j}(v,c_j,c_{-j})\big)\\
  &\hspace{12em}-\big(p_j(v,c_j',c_{-j})-c_j\cdot x_{E_j}(v,c_j',c_{-j})\big)\Big],
  \end{aligned}
\end{equation}
to be maximized over the allocation rules $x$ with $x(v,c)\in\cX(\cF)$ at every profile, and over the free payments $p$. Since the payments are unrestricted in sign, the supremum is $+\infty$ as soon as one of them carries a nonzero coefficient, and the upper bound on $\SB$ that the dual then gives is vacuous.

\begin{lemma}[Useful duals]\label[lemma]{lem:useful}
Let $(\tau,\alpha,\beta)\ge0$. If $(\tau,\alpha,\beta)$ is a useful dual in the sense of \Cref{def:useful}, the supremum of the Lagrangian in \Cref{eq:lagrangian} is finite; otherwise it is $+\infty$.
\end{lemma}

\begin{proof}
Fix a buyer $i$ and a profile $(v,c)$, and let $\rho>0$ be the probability of the other agents' types $(v_{-i},c)$, so that the profile has probability $f_i(v_i)\rho$. The variable $p_i(v,c)$ occurs in the budget term with coefficient $\tau f_i(v_i)\rho$; in the terms with true type $v_i$, where $\cU_i(v_i;v_i)$ contributes $-\rho$ each, with total coefficient $-\Out_i(v_i)\rho$; and in the terms with report $v_i$, where $-\cU_i(v_i';v_i)$ contributes $+\rho$ each, with total coefficient $+\In_i(v_i)\rho$. Dividing by $\rho$, the coefficient vanishes at every profile with the same $v_i$ precisely when \Cref{eq:buyer-flow} holds. The null report contributes no variable of its own, $p_i(\bot,v_{-i},c)$ being the constant $0$. For a seller every sign reverses, since $p_j$ enters the seller's utility positively and the budget constraint negatively; the coefficient is $-\tau g_j(c_j)+\Out_j(c_j)-\In_j(c_j)$ times the probability of the other agents' types, and its vanishing is \Cref{eq:seller-flow}. Since the remaining domain is a product of copies of the polytope $\cX(\cF)$, on which $\cL$ is bounded, \lemref[partial Lagrangian duality]{lem:partial-duality} below shows that the supremum is finite exactly when all these coefficients vanish.
\end{proof}

\paratitle{The coefficients of $\cL$}
For a buyer $i$, a type $v_i\in T_i$ and a trade $e\in E_i$, and for a seller $j$, a type $c_j\in T_j$ and a trade $e\in E_j$, define
\begin{align*}
  \Xi^B_{i,e}(v_i)&:=\frac{1}{f_i(v_i)}\Big[v_{i,e}\Out_i(v_i)-\sum_{v_i'\in T_i}\alpha_i(v_i',v_i)\,v'_{i,e}\Big],\\
  \Xi^S_{j,e}(c_j)&:=\frac{1}{g_j(c_j)}\Big[-c_{j,e}\Out_j(c_j)+\sum_{c_j'\in T_j}\beta_j(c_j',c_j)\,c'_{j,e}\Big],
\end{align*}
and let $\Xi(v,c)\in\R^E$ be the vector with entries
\[
  \Xi_e(v,c):=\big(v_{i(e),e}-c_{j(e),e}\big)+\Xi^B_{i(e),e}(v_{i(e)})+\Xi^S_{j(e),e}(c_{j(e)}).
\]
Then, as a function of the allocation and the payments,
\begin{equation}\label{eq:L-split}
  \begin{aligned}
  \cL&=\sum_{(v,c)}\pi(v,c)\,\Xi(v,c)\cdot x(v,c)\\
  &\quad+\sum_i\sum_{(v,c)}\frac{\pi(v,c)}{f_i(v_i)}\big(\tau f_i(v_i)-\Out_i(v_i)+\In_i(v_i)\big)\,p_i(v,c)\\
  &\quad+\sum_j\sum_{(v,c)}\frac{\pi(v,c)}{g_j(c_j)}\big({-\tau} g_j(c_j)+\Out_j(c_j)-\In_j(c_j)\big)\,p_j(v,c).
  \end{aligned}
\end{equation}
The transfer coefficients are the ones just computed; the transfers at the null report are constants, not variables, and contribute nothing. For the allocation, writing $i=i(e)$ and $j=j(e)$, the variable $x_e(v,c)$ has coefficient $v_{i,e}-c_{j,e}$ in $\GFT(x)$ after dividing by the common factor $\pi(v,c)$; it occurs in $\cU_i(v_i;v_i)$ with coefficient $v_{i,e}/f_i(v_i)$ so divided and in $\cU_i(v_i';v_i)$ with coefficient $v'_{i,e}/f_i(v_i)$, and in no other interim utility of buyer $i$, and symmetrically for seller $j$; collecting the multipliers gives $\Xi^B_{i,e}(v_i)$ and $\Xi^S_{j,e}(c_j)$, so the coefficient of $x_e(v,c)$ in $\cL$ is $\pi(v,c)\,\Xi_e(v,c)$.

\begin{lemma}[Partial Lagrangian duality]\label[lemma]{lem:partial-duality}
For $z=(\tau,\alpha,\beta)\ge0$ let $\cL(x,p;z)$ be the Lagrangian in \Cref{eq:lagrangian} and
\[
  \cL^*(z):=\sup\Big\{\cL(x,p;z):\ x(v,c)\in\cX(\cF)\ \text{for every profile},\ p\ \text{free}\Big\}\in\R\cup\{+\infty\}.
\]
\begin{enumerate}
\item[(i)] $\cL^*(z)\ge\mathrm{OPT}(\programsymbol)$ for every $z\ge0$.
\item[(ii)] $\cL^*(z)<\infty$ if and only if the coefficient of every $p_i(v,c)$ and of every $p_j(v,c)$ in $\cL$ vanishes, and in that case $\cL^*(z)=\sum_{(v,c)}\pi(v,c)\max_{X\in\cF}\sum_{e\in X}\Xi_e(v,c)$.
\item[(iii)] There is $z^*\ge0$ with $\cL^*(z^*)=\mathrm{OPT}(\programsymbol)$.
\end{enumerate}
\end{lemma}

\begin{proof}
(i) If $(x,p)$ is feasible for \Cref{prog:second-best}, each constraint expression multiplied by a component of $z$ in $\cL$ is nonnegative, and $z\ge0$, so $\cL(x,p;z)\ge \GFT(x)$; taking the supremum over feasible points gives $\cL^*(z)\ge\mathrm{OPT}(\programsymbol)$.

(ii) All dualized constraints are homogeneous, so $\cL$ is a linear function of $(x,p)$ without constant term, and \Cref{eq:L-split} displays it as a sum over the allocation and the transfer variables separately. Its supremum over a free transfer variable is finite if and only if that variable's coefficient vanishes; its supremum over $x(v,c)\in\cX(\cF)$ is $\pi(v,c)\max_{X\in\cF}\sum_{e\in X}\Xi_e(v,c)$ and is attained at an indicator $\1_X$, $\cX(\cF)$ being the convex hull of these vectors; and the domain is the product of these, so the supremum is the sum of the terms.

(iii) \Cref{prog:second-best} is a \lemref[finite linear program]{lem:auxlp} with a finite value. Write $u:=(x,p)$ and let $\mathcal D:=\{u:\ x(v,c)\in\cX(\cF)\ \text{for every profile}\}$, a polyhedron because $\cX(\cF)$ is a polytope, and nonempty because $\varnothing\in\cF$ makes $0=\1_\varnothing\in\cX(\cF)$. Let $\Gamma(u)\ge0$ collect the incentive, participation and budget constraints, all linear and homogeneous, so that $\cL(u;z)=\GFT(x)+z\cdot\Gamma(u)$ and $\mathrm{OPT}(\programsymbol)=\max\{\GFT(x):u\in\mathcal D,\ \Gamma(u)\ge0\}$. Fix a description of $\mathcal D$ by finitely many linear inequalities and take the linear-programming dual of \Cref{prog:second-best} written with those inequalities: it is a minimum over $z\ge0$ and the multipliers $\mu$ of the inequalities describing $\mathcal D$, and for fixed $z$ the minimization over $\mu$ is, again by linear-programming duality and because $\mathcal D\ne\varnothing$, the value of $\sup\{\GFT(x)+z\cdot\Gamma(u):u\in\mathcal D\}=\cL^*(z)$, the two being $+\infty$ together. So the value of the dual of \Cref{prog:second-best} is $\inf_{z\ge0}\cL^*(z)$, and it equals $\mathrm{OPT}(\programsymbol)$ by the duality theorem, the dual optimum being attained at some pair $(z^*,\mu^*)$. Since $\mu^*$ is feasible for the inner minimization at $z^*$ we get $\cL^*(z^*)\le\inf_{z\ge0}\cL^*(z)$, and the reverse inequality holds because the right-hand side is an infimum; so the infimum is attained at $z^*$.
\end{proof}

\begin{lemma}[Lagrangian at a useful dual]\label[lemma]{lem:lagrangian}
At a useful dual the coefficient of every payment variable in \Cref{eq:lagrangian} vanishes, and
\begin{equation}\label{eq:lagrangian-value}
  \cL(x,p;\tau,\alpha,\beta)=(1+\tau)\,\E_{v,c}\Big[\sum_{e\in E}\big(\tv_{i(e),e}(v_{i(e)},\alpha,\tau)-\tc_{j(e),e}(c_{j(e)},\beta,\tau)\big)\,x_e(v,c)\Big].
\end{equation}
\end{lemma}

\begin{proof}
At a useful dual the payment coefficients in \Cref{eq:L-split} vanish by \Crefrange{eq:buyer-flow}{eq:seller-flow}. For the allocation, fix a buyer $i$, a type $v_i\in T_i$ and a trade $e\in E_i$. By \Cref{eq:buyer-flow}, $\Out_i(v_i)=\tau f_i(v_i)+\In_i(v_i)$, and by \Cref{eq:transformed}, $\sum_{v_i'\in T_i}\alpha_i(v_i',v_i)(v_{i,e}-v'_{i,e})=(1+\tau)f_i(v_i)\big(\tv_{i,e}(v_i)-v_{i,e}\big)$; hence
\[
  f_i(v_i)\,\Xi^B_{i,e}(v_i)=\tau f_i(v_i)v_{i,e}+\sum_{v_i'\in T_i}\alpha_i(v_i',v_i)(v_{i,e}-v'_{i,e})=f_i(v_i)\big[(1+\tau)\tv_{i,e}(v_i)-v_{i,e}\big].
\]
In the same way, $\Xi^S_{j,e}(c_j)=c_{j,e}-(1+\tau)\tc_{j,e}(c_j)$ for a seller $j$, a type $c_j\in T_j$ and a trade $e\in E_j$. Adding $v_{i(e),e}-c_{j(e),e}$ gives $\Xi_e(v,c)=(1+\tau)\big(\tv_{i(e),e}(v_{i(e)})-\tc_{j(e),e}(c_{j(e)})\big)$ for every $e\in E$, and substituting into \Cref{eq:L-split} gives \Cref{eq:lagrangian-value}.
\end{proof}

\begin{restatedstatement}{Theorem}{prop:dual}{Duality}
At every useful dual $(\tau,\alpha,\beta)$,
\begin{equation*}
  \SB\le (1+\tau)\,\E_{v,c}\big[\tH(v,c,\alpha,\beta,\tau)\big],
  \tag{\ref*{eq:weak-duality}}
\end{equation*}
and the second best is the least of these bounds,
\begin{equation*}
  \SB=\min_{(\tau,\alpha,\beta)\ \text{useful}}\ (1+\tau)\,\E_{v,c}\big[\tH(v,c,\alpha,\beta,\tau)\big],
  \tag{\ref*{eq:sb-min}}
\end{equation*}
the minimum being attained.
\end{restatedstatement}

\begin{proof}[Proof of \thmref{prop:dual}]
Since \Cref{prog:second-best} is a \lemref[finite linear program]{lem:auxlp} expressing the \lemref[second best as a linear program]{lem:sb-is-lp}, it has the finite value $\mathrm{OPT}(\programsymbol)=\SB$. By \lempartref[partial Lagrangian duality]{lem:partial-duality}{(ii)} and \Cref{eq:L-split}, $\cL^*(z)<\infty$ exactly when $z$ is a useful dual, and then, by the expression for $\Xi_e$ in the proof of the \lemref[Lagrangian at a useful dual]{lem:lagrangian},
\[
  \cL^*(z)=\sum_{(v,c)}\pi(v,c)\max_{X\in\cF}\sum_{e\in X}\Xi_e(v,c)=(1+\tau)\,\E_{v,c}\big[\tH(v,c,\alpha,\beta,\tau)\big].
\]
\lempartref{lem:partial-duality}{(i)} therefore gives \Cref{eq:weak-duality}. By \lempartref[partial Lagrangian duality]{lem:partial-duality}{(iii)} some $z^*\ge0$ has $\cL^*(z^*)=\SB<\infty$, so $z^*$ is a useful dual at which \Cref{eq:weak-duality} holds with equality; together with \Cref{eq:weak-duality} this is \Cref{eq:sb-min}, the minimum being attained at $z^*$.
\end{proof}

The set of useful duals is not compact. For a buyer with $|T_i|\ge2$, adding the same $\gamma\ge0$ to $\alpha_i(v_i,v_i')$ for every pair of distinct types adds $\gamma(|T_i|-1)$ to both $\Out_i(v_i)$ and $\In_i(v_i)$ at every type, so \Cref{eq:buyer-flow} continues to hold for every $\gamma\ge0$. Attainment of the minimum in \thmref[duality]{prop:dual} therefore cannot come from compactness; it comes from linear-programming duality.

\clearpage
\subsection{Markov-chain recovery and prior-density bounds}
\label[appendix]{app:markov}

\begin{restatedstatement}{Lemma}{lem:markov}{Markov-chain recovery}
For every buyer $i$, the matrix $\Id-\cA_i$ is invertible, and
\begin{equation*}
  (\Id-\cA_i)^{-1}=\int_0^\infty e^{-\theta}e^{\theta\cA_i}\,\mathrm{d}\theta .
  \tag{\ref*{eq:resolvent}}
\end{equation*}
The matrices $e^{\theta\cA_i}$, $\theta\ge0$, are the transition matrices of a continuous-time Markov chain on $T_i$: for every $\theta\ge0$, the row $e^{\theta\cA_i}(v_i,\cdot)$ is a probability distribution on $T_i$, the law of the state of the chain at time $\theta$ when it starts at $v_i$. By \Cref{eq:resolvent}, if $\theta$ is drawn from $\Exp(1)$, independently of the chain, the state $\hat v_i$ at time $\theta$ has law $(\Id-\cA_i)^{-1}(v_i,\cdot)$. Consequently, the virtual value at this state equals the true value on average:
\begin{equation*}
  \mathop{\E}\limits_{\substack{\theta\sim\Exp(1)\\ \hat v_i\sim e^{\theta\cA_i}(v_i,\cdot)}}\big[\tv_i(\hat v_i,\alpha,\tau)\big]=v_i .
  \tag{\ref*{eq:recovery}}
\end{equation*}
The same holds for every seller $j$, with $\cB_j$, $c_j$ and $\tc_j$ in place of $\cA_i$, $v_i$ and $\tv_i$.
\end{restatedstatement}

\begin{proof}[Proof of \lemref{lem:markov}]
By \Cref{eq:reverse-rates} the off-diagonal entries of $\cA_i$ are nonnegative and its rows sum to zero. For such a matrix, $e^{\theta\cA_i}$ has nonnegative entries and rows summing to one for every $\theta\ge0$, and these are the transition matrices of a continuous-time Markov chain on $T_i$ \citep*[Chapter~2]{Norris1997}.\footnote{For a self-contained argument, choose $\gamma>0$ at least as large as every $-\cA_i(v_i,v_i)$; then $\Pi:=\Id+\cA_i/\gamma$ is a stochastic matrix, $\cA_i=\gamma(\Pi-\Id)$, and $e^{\theta\cA_i}=e^{-\gamma\theta}\sum_{n\ge0}\frac{(\gamma\theta)^n}{n!}\Pi^n$ is a Poisson mixture of the stochastic matrices $\Pi^n$.} Its entries therefore lie in $[0,1]$, so the integral in \Cref{eq:resolvent} converges entrywise and $e^{-\theta}e^{\theta\cA_i}\to0$ as $\theta\to\infty$. Differentiating the power series termwise gives $\frac{\mathrm{d}}{\mathrm{d}\theta}e^{\theta\cA_i}=\cA_ie^{\theta\cA_i}$, hence
\[
  \frac{\mathrm{d}}{\mathrm{d}\theta}\big(e^{-\theta}e^{\theta\cA_i}\big)=-(\Id-\cA_i)\,e^{-\theta}e^{\theta\cA_i},
\]
and integrating over $[0,\infty)$, with the boundary values $\Id$ at $\theta=0$ and $0$ at infinity, yields $(\Id-\cA_i)\int_0^\infty e^{-\theta}e^{\theta\cA_i}\,\mathrm{d}\theta=\Id$. A square matrix with a right inverse is invertible and the right inverse is its inverse, which proves \Cref{eq:resolvent}. Because the rate-one exponential law has density $e^{-\theta}$, the state $\hat v_i$ equals $v_i'\in T_i$ with probability
\[
  \int_0^\infty e^{-\theta}\,e^{\theta\cA_i}(v_i,v_i')\,\mathrm{d}\theta=(\Id-\cA_i)^{-1}(v_i,v_i').
\]
Averaging the rows of $\widetilde V_i$ against this law and using \Cref{eq:transformed-Q},
\[
  \E\big[\tv_i(\hat v_i,\alpha,\tau)\big]=\big((\Id-\cA_i)^{-1}\widetilde V_i\big)(v_i,\cdot)=\big((\Id-\cA_i)^{-1}(\Id-\cA_i)V_i\big)(v_i,\cdot)=v_i ,
\]
which is \Cref{eq:recovery}. The seller statement is proved in the same way.
\end{proof}

\begin{restatedstatement}{Lemma}{lem:density}{Prior-density bounds}
At every useful dual and for every $\theta\ge0$,
\begin{equation*}
    f_i e^{\theta\cA_i}\le e^{\kappa\theta}f_i\quad(i\in I),
    \qquad
    g_j e^{\theta\cB_j}\le e^{\kappa\theta}g_j\quad(j\in J),
    \tag{\ref*{eq:density}}
\end{equation*}
where $\kappa=\tau/(1+\tau)$.
In particular, averaging over the exponential clock gives
\[
    f_i(\Id-\cA_i)^{-1}\le (1+\tau)f_i \quad(i\in I),
    \qquad
    g_j(\Id-\cB_j)^{-1}\le (1+\tau)g_j \quad(j\in J).
\]
\end{restatedstatement}
\begin{proof}[Proof of \lemref{lem:density}]
Fix a buyer $i$, and write $f=f_i$, $A=\cA_i$, and
$\kappa=\tau/(1+\tau)$. For $\theta\ge 0$, let
$p_\theta=f e^{\theta A}$ be the distribution of the chain at time
$\theta$ when it starts from the prior $f$.
Since the generator $A$ of the Markov chain is induced by a flow, \Cref{eq:cap} gives $fA\le\kappa f$ componentwise. Since
$e^{\theta A}$ is entrywise nonnegative, it follows that
\[
    \frac{\mathrm{d}}{\mathrm{d}\theta}p_\theta
    =(fA)e^{\theta A}
    \le \kappa f e^{\theta A}
    =\kappa p_\theta.
\]
For every $v\in T_i$ and $\theta\ge 0$, we have
$p_\theta(v)>0$\footnote{For sufficiently large $\gamma$,
$B=A+\gamma\Id\ge0$ entrywise, so the exponential series gives
$e^{\gamma\theta}\Id e^{\theta A}=e^{\theta B}=\Id+\sum_{n\ge1}\frac{\theta^nB^n}{n!}\ge\Id$ entrywise. Hence $e^{\theta A}\ge e^{-\gamma\theta}\Id$ entrywise, and
$p_\theta(v)\ge e^{-\gamma\theta}f(v)>0$.}
because the chain can start at $v$ and make no jump before time
$\theta$. We may therefore divide by $p_\theta(v)$ and integrate:
\[
    \log\frac{p_\theta(v)}{f(v)}
    =\int_0^\theta
       \frac{1}{p_\vartheta(v)}\frac{\mathrm{d}}{\mathrm{d}\vartheta}p_\vartheta(v)\,\mathrm{d}\vartheta
    \le \kappa\theta.
\]
Exponentiating yields
\[
    f e^{\theta A}\le e^{\kappa\theta}f
\]
componentwise, proving the buyer's prior-density bound in \Cref{eq:density}.

Using the integral representation of $(\Id-A)^{-1}$ and integrating
this inequality against $e^{-\theta}\,\mathrm{d}\theta$, we obtain
\[
    f(\Id-A)^{-1}
    =\int_0^\infty e^{-\theta}f e^{\theta A}\,\mathrm{d}\theta
    \le f\int_0^\infty e^{-(1-\kappa)\theta}\,\mathrm{d}\theta
    =\frac{f}{1-\kappa}
    =(1+\tau)f.
\]
Similarly, we can derive the same guarantees for every seller.
\end{proof}

\clearpage
\section{Special cases of our model}\label[appendix]{sec:examples}
\begin{example}[Bilateral trade \citep*{MyersonSatterthwaite1983,DengMaoSivanWang2022,Fei2022,LiuQinRenWang2026}]\namedstatementlabel{example}{ex:bilateral}{Bilateral trade}
Our model covers bilateral trade by taking one buyer, one seller and a single trade $e$ between them, so that $\cF=\{\varnothing,\{e\}\}$. Each type vector then has a single coordinate, so the types are scalars.
\end{example}

\begin{example}[Matching markets \citep*{BrustleCaiWuZhao2017,BabaioffCaiGonczarowskiZhao2018,BabaioffRubinsteinTanWang2026}]\namedstatementlabel{example}{ex:matching-model}{Matching markets}
A matching market has unit-demand buyers and unit-supply sellers, each holding one private number: buyer $i$'s value for an item, seller $j$'s cost of parting with its own. A downward-closed family of feasible matchings between $I$ and $J$ limits which pairs may trade. Our model recovers this market as the special case in which $E$ is the set of admissible pairs, $\cF$ is the family of feasible matchings, and every coordinate of an agent's type vector equals the agent's value/cost, so each type distribution is supported on the diagonal $\{(t,\dots,t)\}$. Since every feasible set is a matching, no agent executes more than one trade, and additivity over executed trades therefore reduces exactly to unit demand and unit supply.
\end{example}

\begin{example}[One buyer, heterogeneous items \citep*{CaiGoldnerMaZhao2021}]\namedstatementlabel{example}{ex:multi}{One buyer, heterogeneous items}
One buyer faces $n$ sellers, with seller $j$ owning one item; the trades are $E=\{e_1,\dots,e_n\}$ with $e_j$ the sale of item $j$. The buyer is additive over the items it receives, and a public downward-closed $\cF\subseteq2^E$ limits which sets of items it may take. A constrained-additive buyer is exactly an additive buyer restricted to the feasible sets of such a family. Our model is more general even here: \citet*{CaiGoldnerMaZhao2021} require the buyer's values $v_{i,e_1},\dots,v_{i,e_n}$ for the $n$ items to be independent of one another, whereas we allow them to be arbitrarily correlated, requiring independence only across agents.
\end{example}

\begin{example}[Bilateral bargaining]
\label[example]{ex:bargaining}
Two agents with independent private types bargain over a finite
set $A$ of agreements, including a disagreement outcome $a_0$.
Their utilities $u_i(a;\theta_i)$ may be arbitrary; normalize
$u_i(a_0;\theta_i)=0$.

Represent each agreement $a\ne a_0$ by a trade $e_a$ between
one buyer and one seller, and permit at most one trade:
\[
  \cF=\{\varnothing\}\cup
  \big\{\{e_a\}:a\in A\setminus\{a_0\}\big\}.
\]
For a sufficiently large public constant $K$, set
\[
  v_a(\theta_1)=K+u_1(a;\theta_1),
  \qquad
  c_a(\theta_2)=K-u_2(a;\theta_2).
\]
These values and costs are nonnegative, and the gain from
executing $e_a$ equals the total utility gain from agreement $a$.
\end{example}

\end{document}